\documentclass[
 amsmath,
 amssymb,
]{aastex701}
\usepackage{natbib}
\usepackage{graphicx} 
\usepackage{dcolumn} 
\usepackage{algorithm} 
\usepackage{algpseudocode} 

\usepackage{savesym}
\savesymbol{tablenum}
\usepackage{siunitx}
\usepackage{amsmath}
\usepackage{xspace}

\restoresymbol{SIX}{tablenum}

\newcolumntype{d}[1]{D{.}{.}{#1}}
\newcommand{\pluseq}{\mathrel{+}=}

\begin{document}

\def\simlt{\lower.5ex\hbox{$\; \buildrel < \over \sim \;$}}
\def\simgt{\lower.5ex\hbox{$\; \buildrel > \over \sim \;$}}
\ifx \change \undefined
  \newcommand{\change}[1]{{#1}}
\fi
\ifx \forcemathnospace \undefined
  \newcommand{\forcemathnospace}[1]{\ifmmode{\rm{#1}}\else {$\rm{#1}$}\fi\xspace}
\fi
\ifx \forcemath \undefined
  \newcommand{\forcemath}[1]{\ifmmode{\rm{#1}}\else {$\rm{#1}$}\fi\xspace}
\fi
\ifx \unitnospace \undefined
  \newcommand{\unitnospace}[1]{\forcemath{\rm #1}}
\fi
\ifx \unit \undefined
  \newcommand{\unit}[1]{\forcemath{\rm #1}}
\fi
\newcommand{\Aeff}{\forcemath{A_{\rm eff}}}
\newcommand{\Smin}{\forcemath{S_{\rm min}}}
\newcommand{\Emin}{\forcemath{E_{\rm min}}}
\newcommand{\ICON}{{\em ICON}}
\newcommand{\EUV}{{\em ICON EUV}}
\newcommand{\NIST}{{\rm NIST}}
\newcommand{\IUE}{{\em IUE}}
\newcommand{\HST}{{\em HST}}
\newcommand{\IRAS}{{\em IRAS}}
\newcommand{\Spitzer}{{\em Spitzer}}
\newcommand{\FUSE}{{\em FUSE}}
\newcommand{\COBE}{{\em COBE}}
\newcommand{\WMAP}{{\em WMAP}}
\newcommand{\WHAM}{{\em WHAM}}
\newcommand{\VTSS}{{\em VTSS}}
\newcommand{\SHASSA}{{\em SHASSA}}
\newcommand{\ROSAT}{{\em ROSAT}}
\newcommand{\WIRE}{{\em WIRE}}
\newcommand{\SPEAR}{{\em SPEAR}}
\newcommand{\Tycho}{{\em Tycho}}
\newcommand{\Hipparcos}{{\em Hipparcos}}
\newcommand{\GALEX}{{\em GALEX}}
\newcommand{\SETHI}{{{\rm SETH}{\sc i}}}
\newcommand{\CO}{{CO}}
\ifx \degree \undefined
  \newcommand{\degree}{\ifmmode {^{\circ}} \else {$^{\circ}$} \fi}
\fi
\newcommand{\degrees}{\ifmmode {^{\circ}} \else {$^{\circ}$} \fi}
\newcommand{\ie}{{\em i.e.}}
\newcommand{\quarter}{\ifmmode {\frac{1}{4}} \else {$\frac{1}{4}$} \fi}
\newcommand{\angstroms}{\unit{\AA}}
\ifx \km \undefined
  \newcommand{\cm}{\unit{cm}}
\else
  \renewcommand{\cm}{\unit{\cm}}
\fi
\newcommand{\cmpersecsqr}{\unit{\cm\,\s^{-2}}}
\newcommand{\etal}{{\rm et al.}}
\newcommand{\emunits}{\unit{{\rm cm}^{-6} {\rm pc}}}
\newcommand{\cmpersec}{\unit{cm\ s^{-1}}}
\newcommand{\wattpermsqr}{\unit{W\ m^{-2}}}
\newcommand{\joulepermsqr}{\unit{J\ m^{-2}}}
\newcommand{\hzpersec}{\unit{Hz\ s^{-1}}}
\newcommand{\chisqr}{\unitnospace{\chi^2}}
\ifx \km \undefined
  \newcommand{\km}{\unit{km}}
\else 
  \renewcommand{\km}{\unit{km}}
\fi
\newcommand{\rad}{\unit{radians}}
\newcommand{\pressure}{\unit{{\rm cm}^{-3} {\rm K}}}
\ifx \s \undefined
  \newcommand{\s}{\unit{s}}
\else
  \renewcommand{\s}{\unit{s}}
\fi
\newcommand{\seconds}{\s}
\newcommand{\persec}{\unit{s^{-1}}}
\newcommand{\hours}{\hour}
\ifx \us \undefined
  \newcommand{\us}{\unit{\mu s}}
\else
  \renewcommand{\us}{\unit{\mu s}}
\fi
\newcommand{\jyus}{\unit{Jy\ \mu s}}
\newcommand{\Kjy}{\unit{K\ Jy^{-1}}}
\ifx \Hz \undefined
  \newcommand{\Hz}{\unit{Hz}}
\else 
  \renewcommand{\Hz}{\unit{Hz}}
\fi
\ifx \kHz \undefined
  \newcommand{\kHz}{\unit{kHz}}
\else
  \renewcommand{\kHz}{\unit{kHz}}
\fi
\ifx \MHz \undefined
  \newcommand{\MHz}{\unit{MHz}}
\else
  \renewcommand{\MHz}{\unit{MHz}}
\fi
\ifx \GHz \undefined
  \newcommand{\GHz}{\unit{GHz}}
\else
  \renewcommand{\GHz}{\unit{GHz}}
\fi
\newcommand{\obar}[1]{\overline{#1}}
\newcommand{\AU}{\unit{A.U.}}
\newcommand{\lt}{\unit{<}}
\newcommand{\gt}{\unit{>}}
\newcommand{\yr}{\unit{year}}
\newcommand{\tten}[1]{\ifmmode {\times 10^{#1}} \else {$\times 10^{#1}$} \fi}
\newcommand{\tentothe}[1]{\ifmmode {10^{#1}} \else {$10^{#1}$} \fi}
\newcommand{\parsec}{\unit{pc}}
\newcommand{\pcyr}{\unit{pc^{-3}yr^{-1}}}
\newcommand{\kpc}{\unit{kpc}}
\newcommand{\microgauss}{\unit{\mu G}}
\newcommand{\deriv}[2]{\frac{d#1}{d#2}}
\newcommand{\pderiv}[2]{\frac{\partial #1}{\partial #2}}
\newcommand{\del}{\nabla}
\newcommand{\pu}{\unit{ph\ s^{-1}}\unit{cm^{-2}\ str^{-1}}}
\newcommand{\intensity}{\unit{erg\ s^{-1}\ cm^{-2}\ str^{-1}}}
\newcommand{\DM}{\unit{DM}}
\ifx \dm \undefined
  \newcommand{\dm}{\unit{pc\ cm^{-3}}}
\fi
\newcommand{\cu}{\unit{ph\ s^{-1}}\-\unit{cm^{-2}}\- \unit{str^{-1}\- \angstrom^{-1}}}
\newcommand{\HI}{{\rm H{\sc i}}}
\newcommand{\NaI}{{\rm Na{\sc i}}}
\newcommand{\HII}{{\rm H{\sc ii}}}
\newcommand{\Htwo}{\unit{H_2}}
\newcommand{\HeI}{{\rm He{\sc i}}}
\newcommand{\OI}{{\rm O{\sc i}}}
\newcommand{\OII}{{\rm O{\sc ii}}}
\newcommand{\Oplus}{{\rm O$^+$}}
\newcommand{\ArI}{{\rm Ar{\sc i}}}
\newcommand{\CII}{{\rm C{\sc ii}}}
\newcommand{\CIIstar}{{\rm C{\sc ii}$^*$}}
\newcommand{\AlII}{{\rm Al{\sc ii}}}
\newcommand{\SiII}{{\rm Si{\sc ii}}}
\newcommand{\SiIIstar}{{\rm Si{\sc ii}$^*$}}
\newcommand{\MgII}{{\rm Mg{\sc ii}}}
\newcommand{\CIII}{{\rm C{\sc iii}}}
\newcommand{\NIII}{{\rm N{\sc iii}}}
\newcommand{\SiIII}{{\rm Si{\sc iii}}}
\newcommand{\OIII}{{\rm O{\sc iii}}}
\newcommand{\CIV}{{\rm C{\sc iv}}}
\newcommand{\NII}{{\rm N{\sc ii}}}
\newcommand{\SII}{{\rm S{\sc ii}}}
\newcommand{\SIV}{{\rm S{\sc iv}}}
\newcommand{\SiIV}{{\rm Si{\sc iv}}}
\newcommand{\NIV}{{\rm N{\sc iv}}}
\newcommand{\OIV}{{\rm O{\sc iv}}}
\newcommand{\iovi}{{\rm \unit{I_\OVI}}}
\newcommand{\iciv}{{\rm \unit{I_\CIV}}}
\newcommand{\CV}{{\rm C{\sc v}}}
\newcommand{\NV}{{\rm N{\sc v}}}
\newcommand{\SV}{{\rm S{\sc v}}}
\newcommand{\OV}{{\rm O{\sc v}}}
\newcommand{\CVI}{{\rm C{\sc vi}}}
\newcommand{\NVI}{{\rm N{\sc vi}}}
\newcommand{\OVI}{{\rm O{\sc vi}}}
\newcommand{\NeVI}{{\rm Ne{\sc vi}}}
\newcommand{\ArVI}{{\rm Ar{\sc vi}}}
\newcommand{\doublet}{\ifmmode {\lambda\lambda} \else {$\lambda\lambda$} \fi}
\newcommand{\singlet}{\ifmmode {\lambda} \else {$\lambda$} \fi}
\newcommand{\cmsquared}{\unit{cm^2}}
\newcommand{\cmcubed}{\unit{cm^3}}
\newcommand{\kmpersec}{\km\persec}
\newcommand{\kmpersecond}{\kmpersec}
\newcommand{\percmcubed}{\unit{cm^{-3}}}
\newcommand{\cmsqr}{\unit{cm^2}}
\newcommand{\percmsqr}{\unit{cm^{-2}}}
\newcommand{\Halpha}{\unit{H\alpha}}
\newcommand{\Lyb}{\unit{Ly\,\beta}}
\newcommand{\LyB}{\unit{Ly\,\beta}}
\newcommand{\Lya}{\unit{Ly\,\alpha}}
\newcommand{\LyA}{\unit{Ly\,\alpha}}
\newcommand{\crateflux}{\unit{s^{-1} cm^{-2}}}
\newcommand{\eps}{\varepsilon}
\newcommand{\ksps}{\unit{{\rm ksps}}}
\newcommand{\Msps}{\unit{{\rm Msps}}}
\ifx \jahh \undefined
  \newcommand{\jahh}{{\rm JAHH}}
\fi
\ifx \jgr \undefined 
  \newcommand{\jgr}{{J.\,Geophys.\,Res.\xspace}}
\fi
\ifx \pasp \undefined
  \newcommand{\pasp}{{Pub.\,Astron.\,Soc.\,Pac.\xspace}}
\fi
\ifx \jgrsp \undefined
  \newcommand{\jgrsp}{{J.\,Geophys.\,Res.\,(Space\,Phys.)\xspace}}
\fi
\ifx \aspc \undefined
  \newcommand{\aspc}{{ASP\,Conf.\,Ser.\xspace}}
\fi
\ifx \iauc \undefined
  \newcommand{\iauc}{{IAU~Colloq.}}
\fi
\ifx \grl \undefined
  \newcommand{\grl}{{Geophys.\,Res.\,Let.\xspace}}
\fi
\ifx \solphys \undefined
\newcommand{\solphys}{{Sol.\,Phys.\xspace}}
\fi
\ifx \aj \undefined
  \newcommand{\aj}{{Astron.\,J.\xspace}}
\fi
\ifx \apj \undefined
  \newcommand{\apj}{{Astrophys.\,J.\xspace}}
\fi
\ifx \apjl \undefined
  \newcommand{\apjl}{{Astrophys.\,J.\xspace}}
\fi
\ifx \apjs \undefined
  \newcommand{\apjs}{{Astrophys.\,J.\,Suppl.\xspace}}
\fi
\ifx \aap \undefined
  \newcommand{\aap}{{Astron.\,&\,Astrophys.\xspace}}
\fi
\ifx \aas \undefined
  \newcommand{\aas}{{AAS Meeting Abstracts\xspace}}
\fi
\ifx \procspie \undefined 
  \newcommand{\procspie}{{Proc.\,SPIE}}
\fi
\ifx \spiec \undefined
  \newcommand{\spiec}{\procspie}
\fi
\ifx \ieeeproc \undefined
  \newcommand{\procieee}{IEEE\,Proc.\xspace}
\fi
\ifx \ieeeproc \undefined
  \newcommand{\ieeeproc}{IEEE\,Proc.\xspace}
\fi
\ifx \jgridcomp \undefined
  \newcommand{\jgridcomp}{{J.\,Grid\,Comput.\xspace}}
\fi
\ifx \revsciinst \undefined
  \newcommand{\revsciinst}{{Rev.\,Sci.\,Instr.\xspace}}
\fi
\ifx \skytel \undefined
  \newcommand{\skytel}{{Sky\,\&\,Telescope}}
\fi
\ifx \jai \undefined
  \newcommand{\jai}{{\rm J. Astron. Instr.}}
\fi
\ifx \areps \undefined
  \newcommand{\areps}{{\rm Annu. Rev. Earth \& Planet. Sci.}}
\fi
\ifx \detokenize \undefined
  \newcommand{\detokenize}[1]{{#1}}
\fi
\ifx \doi \undefined
  \newcommand{\doi}[1]{DOI \href{http://dx.doi.org/\detokenize{#1}}{\ttfamily \detokenize{#1}}\spc}
\fi
\ifx \cise \undefined
  \newcommand{\cise}{{Computi. Sci. \& Eng.}}
\fi

\newcommand{\var}[1]{{\ifmmode {{#1}}\else {${#1}$}\fi}\xspace}

\newcommand{\sahtotalbw}{\var{\Delta\nu_{\rm tot}}}
\newcommand{\sahsamplerate}{\var{\Delta\nu_{\rm wu}}}
\newcommand{\sahwuduration}{\var{\Delta t_{\rm wu}}}
\newcommand{\AObeamwidth}{\var{\theta_{\rm beam}}}
\newcommand{\AOdoppler}{\var{\Delta\nu_{\rm AO}}}
\newcommand{\AOchirp}{\var{\frac{d(\AOdoppler)}{dt}}}
\newcommand{\Dtime}{\var{t(D)}}
\newcommand{\Dbeam}{\var{b(D)}}
\newcommand{\Dpos}{\var{pos(D)}}
\newcommand{\Ddur}{\var{\tau(D)}}
\newcommand{\Dscore}{\var{S(D)}}
\newcommand{\Dsnr}{\var{P(D)}}
\newcommand{\Dfftlen}{\var{\ell(D)}}
\newcommand{\Dnutopo}{\var{\nu_{\rm topo}(D)}}
\newcommand{\nutopo}{\var{\nu_{\rm topo}}}
\newcommand{\Dnubary}{\var{\nu_{\rm bary}(D)}}
\newcommand{\Dnuadj}{\var{\nu_{\rm adj}}}
\newcommand{\Dchirp}{\var{\frac{\Delta\nutopo}{\Delta t}(D)}}
\newcommand{\Dbarychirp}{\var{\frac{\Delta\Dnubary}{\Delta t}(D)}}
\newcommand{\Dperiod}{\var{p(D)}}
\newcommand{\Ddelay}{\var{\delta \tau(D)}}
\newcommand{\Dposprime}{\var{pos(D')}}
\newcommand{\Upos}{\var{\sigma_{pos}}}
\newcommand{\Ufreq}{\var{\sigma_{\nu}}}
\newcommand{\Bpos}{\var{\left(\alpha_{\rm B},\delta_{\rm B}\right)}}
\newcommand{\Bfreq}{\var{\nu(B)}}
\newcommand{\Bbandwidth}{\var{\Delta\nu(B)}}
\newcommand{\BSNR}{\var{P(B)}}
\newcommand{\Bflux}{\var{F(B)}}
\newcommand{\Ppos}{\var{pos(P)}}
\newcommand{\Ptime}{\var{t(P)}}
\newcommand{\Sprob}{\var{S_{\rm prob}(M)}}
\newcommand{\Smultipletpower}{\var{S_{\rm power}(M)}}
\newcommand{\Smultipletdensity}{\var{S_{\rm density}(M)}}
\newcommand{\fftlen}{\var{\ell}}
\newcommand{\fftlenres}{\var{\Delta\nu(\ell)}}
\newcommand{\fftlendur}{\var{\Delta t(\ell)}}
\newcommand{\driftdt}{\var{D_{\Delta t}}}
\newcommand{\driftmaxdrift}{\var{D_{drift_{max}}}}
\newcommand{\driftminangle}{\var{D_{\theta_{min}}}}
\newcommand{\driftntriangles}{\var{D_{n}}}
\newcommand{\driftpa}{\var{D_{prob_1}}}
\newcommand{\driftpb}{\var{D_{prob_2}}}
\newcommand{\zonefrac}{\var{Z_{frac}(t, \fftlen)}}
\newcommand{\zonebinsize}{\var{Z_{\Delta \nu}}}
\newcommand{\zonewindow}{\var{Z_{\Delta t}}}
\newcommand{\zoneprob}{\var{Z_{prob}}}
\newcommand{\periodwindow}{\var{P_{\Delta t}}}
\newcommand{\periodprob}{\var{P_{prob}}}
\newcommand{\periodangle}{\var{P_{\Delta \theta}}}
\newcommand{\mdfbary}{\var{M_{\Delta\nu}(bary)}}
\newcommand{\mdfnonbary}{\var{M_{\Delta\nu}(nonbary)}}
\newcommand{\mdtlocal}{\var{M_{\Delta t}(local)}}
\newcommand{\mdflocal}{\var{M_{\Delta \nu}(local)}}
\newcommand{\mdclocal}{\var{M_{\Delta c}(local)}}
\newcommand{\mdpbary}{\var{M_{\Delta p}(bary)}}
\newcommand{\mdpnonbary}{\var{M_{\Delta p}(nonbary)}}

\newcommand{\maxdelay}{\var{\pm 6.7 \s}}
\newcommand{\maxdft}{\var{128Ki\xspace}}
\newcommand{\mindft}{\var{8\xspace}}
\newcommand{\noctaves}{\var{15\xspace}}
\newcommand{\nchirps}{\var{123\,000}\xspace}
\newcommand{\minbin}{\var{0.075\,\Hz}}
\newcommand{\mintimebin}{\var{8.1\tten{-3}\,\s}}
\newcommand{\maxbin}{\var{1221\,\Hz}}
\newcommand{\maxtimebin}{\var{13.4\,\s}}
\newcommand{\minchirp}{\var{-100\,\hzpersec}}
\newcommand{\maxchirp}{\var{100\,\hzpersec}}
\newcommand{\barycentricwindow}{\var{\pm 125\,\Hz}}
\newcommand{\nonbarycentricwindow}{\var{\pm 125\,\kHz}}
\newcommand{\fullband}{\var{2.5\,\MHz}}
\newcommand{\recordersamplerate}{\var{2.5\,\Msps}}
\newcommand{\bandcenter}{\var{1.42\,\GHz}}
\newcommand{\HIfreq}{\var{1.4204 GHz}}
\newcommand{\maxsendershift}{\var{308\,\kHz}}
\newcommand{\chirprange}{\var{\pm 100\,\hzpersec}}
\newcommand{\baryvelrange}{\var{\pm 29.9\,\kmpersec}}
\newcommand{\baryfreqrange}{\var{\pm 142\,\kHz}}
\newcommand{\wusamplerate}{\var{9.766\,\ksps}}
\newcommand{\wubandwidth}{\var{9.766\,\kHz}}
\newcommand{\wusamples}{\var{2^{20}}}
\newcommand{\wuduration}{\var{107.37\,\s}}
\newcommand{\alfanyears}{\var{14\,{\rm years}}}

\setlength{\parskip}{.5em}
\title[SETI@home: Data Analysis and Findings]{ SETI@home: Data Analysis and Findings}

\author{David~P.~Anderson}
\email{davea@berkeley.edu}
\affiliation{ 
Space Sciences Laboratory, University of California, Berkeley, CA
94720-7450
}%
\author{Eric~J.~Korpela}
\email{korpela@berkeley.edu}
\affiliation{ 
Space Sciences Laboratory, University of California, Berkeley, CA
94720-7450
}%

\author{Dan Werthimer}
\affiliation{ 
Space Sciences Laboratory, University of California, Berkeley, CA
94720-7450
}%
\email{danseti@berkeley.edu}

\author{Jeff Cobb}
\affiliation{ 
Space Sciences Laboratory, University of California, Berkeley, CA
94720-7450
}%
\email{jeffcobb@berkeley.edu}

\author{Bruce Allen}
\affiliation{Max Planck Institut f\"{u}r Gravitationsphysik (Albert Einstein Institut), Hanover, Germany}
\email{bruce.allen@aei.mpg.de}

\date{\today}

\begin{abstract}
SETI@home is a radio Search for Extraterrestrial Intelligence (SETI) project that looks for technosignatures
in data recorded at the Arecibo Observatory.
The data were collected over a period of \alfanyears and
cover almost the entire sky visible to the telescope.
The first stage of data analysis found billions of {\em detections}:
brief excesses of \change{continuous or pulsed narrowband} power.
The second stage removed detections that were likely radio frequency interference (RFI),
then identified and ranked {\em signal candidates}:
groups of detections, possibly spread over the \alfanyears,
that plausibly originate from a single cosmic source.
We manually examined the top-ranking signal candidates
and selected a few hundred.
In the third and final stage we are reobserving the corresponding
sky locations and frequency ranges using the Five-hundred-meter Aperture Spherical Telescope (FAST)
radio telescope.
This paper covers SETI@home's second stage of data analysis.
We describe the algorithms used to remove RFI
and to identify and rank signal candidates.
To guide the development of these algorithms,
we used artificial {\em candidate birdies} that model persistent ET signals
with a range of power, bandwidth, and planetary motion parameters.
This approach also allow\change{ed} us to estimate the sensitivity
of our detection system to these signals.
\end{abstract}


\setlength{\parindent}{0ex}
\section{Introduction
\label{section:intro}}

\subsection{Background}

\change{The question of whether life exists in other parts of the universe is
important and unanswered.
The 1952 Muller-Urey experiment \citep{miller1953,miller1959}
demonstrated the possibility of abiotic
production of the molecular components of living systems.
The detection of
amino acids in meteorites \citep{pearce2015}
and prebiotic molecules in interstellar space \citep{zeng2019, rivilla2023} showed that such
processes are possible
even outside a planetary atmosphere.

The direct detection of living organisms
outside the Solar System remains unlikely in the near future.
A more likely scenario is an indirect detection,
such as an atmospheric biosignature: a compound
released into the atmosphere by biological processes.
However, such compounds may also have an abiogenic source,
so it is uncertain whether such a detection indicates life \citep{tokadjian24,court12}.

Detection of intelligence would provide more certain evidence of life.
An extraterrestrial intelligence (ETI) could create artifacts, signals, or processes that are detectable at interstellar
distances and have no natural counterpart.
Such processes could be a form of radiation
(electromagnetic, particle, or gravitational)
or a physical artifact
(a spacecraft or object passing through or remaining in the Solar System, 
a structure detectable at interstellar distance, 
or an atmospheric component that only has
a technological means of production).
Such items are collectively known as
{\em technosignatures} \citep{technosignatures}.

Due to the relative ease of creating and detecting radio waves
and the relative transparency of atmospheres and interstellar space to such waves,
radio has been proposed as a means of detecting extraterrestrial
intelligence \citep{cocconi59}.
Two primary approaches have been used for such searches:
{\em sky surveys} cover a large fraction of the solid angle of the entire sky, and {\em targeted searches} focus on individual stars or galaxies \citep{search_strategy}.

Several targeted searches have been performed,
including OZMA and OZMA II at Green Bank
\citep{drake60, sagan75, drake86, gray21}, 
Phoenix at the Arecibo Observatory
\citep{backus02}
and at the Allen Telescope Array (ATA),
and Breakthrough Listen projects at the Parkes and Green Bank observatories
\citep{price20, enriquez17}. 
Recently, Breakthrough Listen has begun to observe targets at the Very Large Array \citep{tremblay24} and MeerKAT \citep{czech21}.
In addition, observations of multiple targets have been made at
the FAST observatory in China \citep{luan23} and the ATA \citep{tusay24}.

There have also been a number of sky surveys.
Some have operated {\em commensally},
collecting data from a telescope while
its pointing was being controlled by other projects.
These include searches using various
generations of the SERENDIP spectrometer
at the Hat Creek and Green Bank observatories
\citep{werthimer88}
and at the Arecibo observatory
\citep{cobb00, bowyer16}.
Other sky surveys used dedicated telescopes.
These include the early Ohio State
project and its ``Wow!" signal
\citep{kraus77}, the ``Fly's Eye" project \citep{siemion12} and a brief survey of the anti-solar point \citep{hort24}
at the ATA.  

To date, no repeatable detections of interstellar
technosignatures have been made.
}
\subsection{SETI@home}

The SETI@home project, described here, is a commensal sky survey.
\change{We r}ecorded and analyzed data
from the Arecibo radio observatory \change{from 2006 through 2020},
looking for \change{radio technosignatures}.

\begin{figure}[tbp]
\centerline{
\includegraphics[width=12cm]{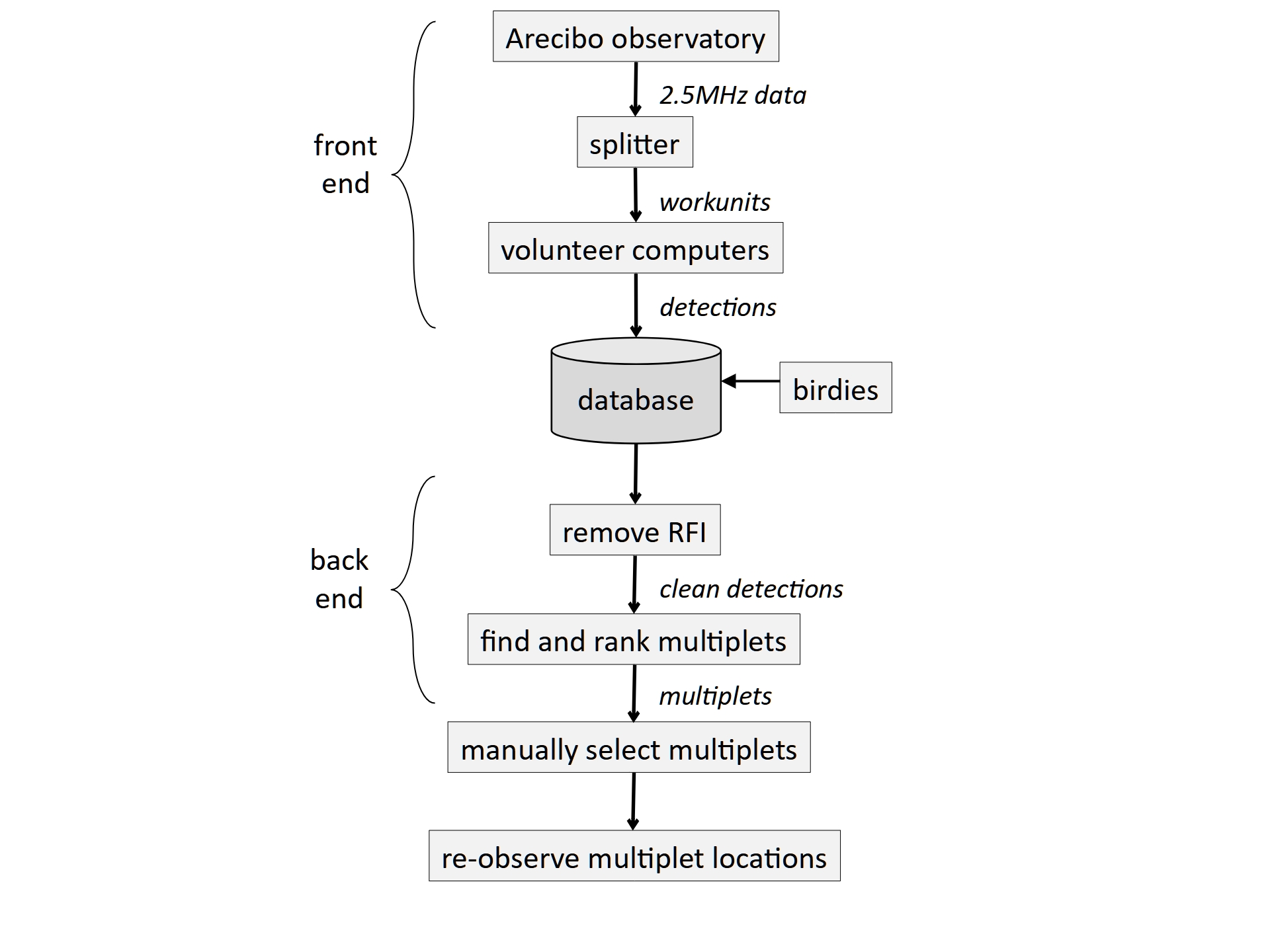}
}
\caption{The SETI@home data analysis process.
\label{fig:pipeline}}
\end{figure}

As shown in Figure~\ref{fig:pipeline},
the {\em front end} of SETI@home digitizes and records the radio signal,
then processes \change{these} data to find {\em detections}:
brief excesses of \change{continuous or pulsed narrowband} power.
This processing is compute-intensive;
it was done using the {\change{Berkeley Open Infrastructure for Network Computing (BOINC)} volunteer computing system \citep{anderson20}
with a pool of several hundred thousand home computers.
The front end of SETI@home is described in detail in \citet{instrument_paper}.

The {\em back end} of SETI@home takes these detections
(about 12 billion of them) as input,
identifies and removes radio frequency interference (RFI),
and finds signal candidates, or {\em multiplets}: groups of detections
whose properties are consistent with those of an ET {\em target signal}.
\change{These detections are selected from the entire duration of the project.
Each multiplet} is assigned {\em scores} \change{inversely proportional to the probability that it is the result of random noise}.

The output of the back end is a set of about 20 million multiplets,
ranked by score.
These were examined, in decreasing score order,
by experts trained to distinguish possible ET signals from noise and RFI.
This process is labor intensive,
and we were able to examine only $\sim 1000$ multiplets.
Thus, we say that we have {\em uncovered} a target signal
if a multiplet that includes detections from the signal appears in
the 1000 top-scoring multiplets.

The manual inspection produced a set of roughly 200 multiplets.
Each of these will be reobserved:
its sky position and frequency range will be examined
with greater sensitivity than was originally available.
In addition, we will control the telescope pointing,
so that each location can be observed longer,
and we can use techniques such as on/off observing.

The back end involves many heuristic algorithms.
We want to ensure that these algorithms work as intended:
in other words, if our data contained a target ET signal, then
a) the detections resulting from the signal would not be rejected as RFI,
b) the back end would find a multiplet composed of these detections,
and c) the multiplet would have sufficiently high scores,
compared with noise multiplets, that it would be uncovered in the above sense.
For this purpose, we inject artificial {\em birdies} into our data,
modeling ET signals with a range of parameters
(e.g. flux, position, planetary motion of the transmitter).
If the back end uncovers these birdies,
we have some confidence that it would uncover a similar ET signal.

The probability that SETI@home uncovers a target ET signal depends on
several factors: most notably the flux of the signal,
its bandwidth,
and the length of our observations of its sky position.
The birdie signal injection mechanism gives us a way to estimate this probability
as a function of these parameters: we can insert many birdies
with various parameters, and see how many we uncover -- i.e.,
what fraction of birdies produce multiplets whose scores
rank in the top 1000 non-birdie multiplets.

This paper describes the SETI@home back end and presents its results.
\S\ref{section:target} describes the types of signals we are looking for.
\S\ref{section:frontend} summarizes the SETI@home front end.
\S\ref{section:coverage} describes the sky coverage of our data.
The birdie mechanism is described in \S\ref{section:birdies}.
Sections~\ref{section:rfi} and~\ref{section:multiplets} describe the algorithms for RFI removal and multiplet detection.
\S\ref{section:implementation} describes the implementation of the back end.
\S\ref{section:results} presents our results.
\S\ref{section:related} discusses related work and the contributions of SETI@home,
and \S\ref{section:future} proposes future work.

\section{Target signals
\label{section:target}}

SETI@home looks for a range of {\em target signals} 
with characteristics typical of technological \change{origin,}
but not known to occur naturally.
Specifically, we look for
a) \change{continuous narrowband} signals whose power is concentrated in a small frequency band
($<$ \maxbin),
b) pulsed \change{narrowband} signals, \change{also} with bandwidths up to about \maxbin,
that turn on and off with a constant but unknown period and duty cycle\change{,
and c) signals having a repeating structure
with periods up to several seconds; see \cite{harp16}.}

Natural sources of \change{such signals} are not known.
The narrowest known natural source of radio emissions (OH masers, \citeauthor{ohmasers} \citeyear{ohmasers}) is about 500 Hz in bandwidth,
\change{but} such masers are not found in the band observed by SETI@home.
The narrowest known emissions in the SETI@home band are about 1 km/s (4750 Hz) in width \change{\citep{peek10,peek11}.}
Similarly, there are no known natural sources of \change{narrowband} pulses.
Pulsars generate \change{broadband} pulses on the \change{timescales} covered by SETI@home,
but these pulses are removed by a filter that eliminates broadband features from the SETI@home analysis \citep{instrument_paper}.

We look for signals that are transmitted over a long period
(ideally our entire observation period) and with a high duty cycle.
It is unlikely that we would detect a transient signal
(one lasting a few seconds or minutes),
because the probability that we would be observing its location
during that period is small.

We assume that the signal transmitter is
in a nearly inertial frame (or Doppler corrected to appear as such),
or is an uncorrected transmitter on the surface of a rotating planet orbiting a star,
in orbit around a planet orbiting a star,
or orbiting a star.

We assume that the movement of the transmitter relative to Earth
is small enough that its sky position as viewed from Earth
is nearly constant during the \alfanyears observing period.
This would not be true for a transmitter located
within the Solar System.

\subsection{Doppler shift}

The frequency at which Arecibo receives a signal
is Doppler-shifted by the velocities of both
transmitter and receiver in the direction of the signal path
at the times of transmission and receipt, respectively.
The shift due to receiver motion,
and the time derivative of this shift, are known.
They are a result of the accelerations of
the Arecibo receiver's reference frame due to
Earth rotation and Earth orbit.

The sender could apply a correction to the transmission frequency
to cancel the changing shift due to the transmitter's acceleration.
If a signal is intended as a beacon and is directional,
it presumably would be adjusted in this way.
We call such a signal {\em barycentric}.
After the correction for receiver motion is applied,
the signal would be received at a nearly constant frequency.

Other \change{potential} ET signals (leakage or omnidirectional beacons)
would not be corrected in this way
and would have a Doppler shift corresponding
to the changing velocity of the transmitter.
We call such signals {\em nonbarycentric}.
After receiver correction,
they would appear to drift in frequency at a varying rate.
The ranges of the sender shift and its derivative
depend on the movements of the transmitter.
We look for target signals for which these
ranges are consistent with the motions
of habitable-zone planets orbiting F and G type stars.
These assumptions are detailed in \S\ref{section:results}.

\section{The SETI@home front end
\label{section:frontend}}

The SETI@home front end is described in a companion paper \cite{instrument_paper}.
We summarize it here.

\subsection{The Arecibo radio telescope}

SETI@home used the radio telescope at the Arecibo Observatory (AO),
a fixed dish with 305-meter aperture.
\change{Its maximum usable viewing angle was 20\degrees away from the zenith.
Given its location at latitude 18\degrees north,
this resulted in a visible area consisting
of a band from -2\degrees to 38\degrees Dec,
covering about 25\% of the sky.
SETI@home can detect signals only within this area \citep{Korpela11a}.
}

Various receivers \change{could} be positioned in \change{the} focal plane.
\change{Our observations began in 1998 using a single-polarization line-feed. \citep{korpela01,korpela02}
In 2006, observations continued using the 7-beam
dual polarization Arecibo L-band Feed Array (ALFA) receiver \citep{cordes06}.
The ALFA receiver
provided significantly better sensitivity ($\gt 10\times$)
and a wider field of view;
it also enabled multibeam RFI rejection (\S\ref{subsection:multibeam}).
Hence, the observations using ALFA superseded the earlier observations.}

ALFA consist\change{ed} of 14 separate receivers,
grouped in polarization pairs, or {\em beams}.
The \change{seven} beams \change{were} arranged in a hexagonal pattern.
The sensitivity of a given receiver can be approximated by a Gaussian
function of the angle from its center.
The half-power width of a beam \change{was} 0\fdg 05;
we denote this $\AObeamwidth$.

The beams point\change{ed} in slightly different directions,
so they are subject to different Doppler shifts.
We denote the shift for beam $b$ at time $t$ by $\AOdoppler(b, t)$,
and its time derivative by $\AOchirp(b, t)$.
The largest component of \AOdoppler\ is typically due to the orbital velocity of Earth.
$\AOchirp$ is dominated by acceleration \change{towards} Earth's rotational axis and is therefore negative, with a value of $\sim -0.16\ \hzpersec$.

\subsection{Data recording and splitting
\label{subsection:data_recording}}

SETI@home covers a frequency band of \change{\fullband} centered at \change{\bandcenter},
near the fine structure transition of the ground state of \HI.
We chose the hydrogen line because it is considered to be
a likely frequency for deliberate beacon transmissions.
ETI who are aware of the \HI\ transition
are likely to use it to \change{study} the structure of the galaxy,
so there are potentially a large number of observers at this frequency.

The SETI@home band is limited to 2.5 MHz
because of performance limitations in recording data
and distributing it over the Internet.
This band is sufficiently large to contain signals near 1.42 GHz,
even after Doppler shifting by receiver motion
and by the target range of transmitter motion.

The analog feed from each of ALFA's 14 receivers
(7 feeds with dual linear polarizations)
goes into a {\em data recorder} computer,
where it is down-converted from 1.42 GHz to quadrature baseband.
A low-pass filter eliminates signals outside the central 2.5 MHz of the band.
The resulting signal is sampled at 2.5 Msps,
with each sample being a 2-bit complex number, one bit real and one bit imaginary. 
The resulting data encode the \change{entire} \fullband frequency band centered
at \bandcenter.

The sky position (RA/Dec) of each ALFA beam is sampled once per second.
These {\em pointing records} are included in the data stream.

The digitized data \change{were} transported from Arecibo to Berkeley on disk drives.
\change{It was} divided using a polyphase filter bank
into 256 frequency {\em subbands} of 9765.625 Hz each.
Each subband was sampled at that rate (with complex samples).
These streams of samples were then divided into segments
of length $2^{20}$ samples = 1 Mi samples (107.37 s in duration).
We call these segments {\em workunits}.
Workunits in a given subband
overlap in time by approximately 20 seconds, so that the
longest features of interest -- 13 seconds or so -- are always contained entirely in at least one workunit.

\subsection{Detections\label{subsection:detections}}

The workunits were analyzed on home computers
using a program that finds detections.
Details of this analysis are given in \citet{instrument_paper};
we briefly summarize it here.

The data were converted to the frequency domain
using the discrete Fourier transform (DFT).
We used \noctaves DFT lengths,
ranging by powers of two from \mindft samples (\maxbin resolution, \mintimebin)
up to \maxdft samples (\maxbin resolution, \maxtimebin).
Long DFTs are sensitive to \change{continuous narrowband} signals;
short DFTs are sensitive to short or rapidly pulsing signals. \change{(See \citeauthor{instrument_paper} \citeyear{instrument_paper} Sec.~5.3 for a discussion of sensitivity.)}
For each DFT length, we computed a sequence of DFTs on the data,
producing a 2-D array of power as a function of time and frequency.

As described earlier, received signals may drift in frequency
due to accelerations of transmitter and receiver.
Rather than looking for features that drift in the power arrays,
such signals can be detected with greater sensitivity
using coherent integration,
in which the data were {\em de-drifted} at a specific Doppler drift rate
before computing DFTs.
This can put drifting features into single frequency bins
where they can be more easily detected.

The program first perform\change{ed} a baseline smoothing operation on the data,
removing any features wider than about 2 KHz.
The data were then de-drifted at a range of Doppler drift rates
corresponding to the range
of planetary motions under consideration
(typically \nchirps rates, ranging from \minchirp to \maxchirp;
see \citet{instrument_paper}).
For each drift rate, the de-drifted data were analyzed
at each of the \noctaves DFT lengths:
we computed the 2-D power array, then looked for features in the array.
We looked for several types of features, or {\em detections}:

\begin{description}
\item[Spike] \change{A short continuous narrowband signal:
a} single DFT bin whose power is at least
24 times the mean noise power.
This threshold was chosen so that workunits
containing Gaussian noise yield an average of 1 spike;
in real data the average is about 7 spikes.
\item[Gaussian] A \change{longer continuous narrowband signal:}
a sequence of DFT bins at the same frequency
whose powers approximate
the Gaussian-shaped envelope that would result from the beam
moving over a fixed source,
given the beam's angular velocity during the workunit.
\item[Pulse] A \change{pulsed narrowband signal:}
the time sequence
of DFT powers at a given frequency approximates a pulsed signal.
These are found using a folding algorithm \citep{staelin69,instrument_paper}.
\item[Triplet] A \change{simple type of pulsed narrowband signal
consisting of a} group of three DFT bins at the same frequency,
above a threshold, and equally spaced in time.
\item[Autocorrelation]
\change{A signal having a repeating structure with periods up to \maxdelay.}
These are found by computing the autocorrelation of the data
\change{with a range of delays,}
looking for delays where the correlation is above a threshold.
This detects any periodic structure in the data.
\change{See \citet{instrument_paper} for more details.}
\end{description}

The number of detections of each type
and their distribution among DFT lengths
is shown in Tables \ref{table:detection_counts}
and \ref{table:detection_counts_dft}.

\begin{table}[tbp]
\begin{center}
\begin{tabular}{l r}
Detection type & Number of detections \\
\hline
Spike & 5,031,283,756 \\
Gaussian & 317,563,087 \\
Pulse & 3,078,135,522 \\
Triplet & 2,561,366,366 \\
Autocorrelation & 1,118,691,234 \\
\end{tabular}
\caption{Detection counts\change{\label{table:detection_counts}}}
\end{center}
\end{table}

\begin{table}[tbp]
\begin{center}
\begin{tabular}{r c c c c c c c}
DFT length & DFT duration & Bandwidth & Spike & Gaussian & Pulse & Triplet & \change{Autocorrelation}\\
samples & {\rm s} & {\rm Hz} & \% &  \% &  \% &  \% & \% \\
\hline
8 & 0.0008 & 1220.7 & & & 6.15 & 7.25 & \\
16 & 0.0016 & 610.3 & & & 1.13 & 10.4 & \\
32 & 0.0032 & 305.1 & 0.001 & & 3.80 & 18.6 & \\
64 & 0.0065 & 152.6 & 0.013 & & 7.33 & 21.8 & \\
128 & 0.0131 & 76.2 & 0.047 & & 12.2 & 18.0 & \\
256 & 0.0262 & 38.1 & 0.091 & & 15.7 & 11.6 & \\
512 & 0.0524 & 19.1 & 0.223 & 0.0 & 16.1 & 6.45 & \\
1024 & 0.104 & 9.53 & 0.391 & 0.003 & 14.9 & 3.07 & \\
2048 & 0.209 & 4.76 & 0.717 & 0.039 & 13.3 & 1.69 & \\
4096 & 0.419 & 2.38 & 1.35 & 0.301 & 7.66 & 0.694 & \\
8192 & 0.838 & 1.19 & 2.82 & 0.796 & 1.46 & 0.092 & \\
16384 & 1.67 & 0.596 & 6.59 & 98.862 & & 0.0 & \\
32758 & 3.35 & 0.298 & 11.5 & & & & \\
65536 & 6.71 & 0.149 & 15.5 & & & & \\
131072 & 13.42 & 0.074 & 60.6 & & & & 100.0 \\
\end{tabular}
\end{center}
\caption{Percentage of detections as a function of DFT length
\label{table:detection_counts_dft}}
\end{table}

We denote the properties of a detection $D$ as follows:
\begin{description}
\item[$\Dtime$] The time midpoint of the DFT bin, or range of bins, where $D$ occurred.
\item[$\Dbeam$] The ALFA beam (0 to 6) in which $D$ was found.
\item[$\Dpos$] The sky position of the beam's center at time $\Dtime$.
This is computed by linearly interpolating between
the two beam pointings before and after $\Dtime$.
\item[$\Ddur$] The duration of the detection.
For spikes, this is the DFT bin duration.
For Gaussians, it is twice the standard deviation of the Gaussian curve.
For pulses and triplets, it is the beam-crossing time
based on the workunit's average beam velocity
(and at most the workunit's duration, 107.37 s).
For autocorrelations it is the duration of
the longest DFT length.
\item[$\Dscore$] An estimate of the probability
of $D$ occurring in noise. \change{For spikes and triplets this is} based on \change{peak} power.
For Gaussians, it\change{'s based on power and} goodness of fit\change{. F}or pulses,
its statistics based on the number of samples added into
each bin of the folded array.
These values are negated, so high scores are better.
See \citet{instrument_paper} for details.
\end{description}

Detection types other than autocorrelation also have the following \change{attributes}:
\begin{description}
\item[$\Dsnr$] The detection power as
a multiple of the mean noise power.
\item[$\Dfftlen$] The DFT length at which $D$ was found.
\item[$\Dnutopo$] The frequency in the reference frame
of the observatory; that is, the frequency of the DFT bin where $D$
 occurred, correcting for de-drifting.
This is used in RFI detection, since terrestrial RFI
isn't Doppler-shifted by receiver movement.
\item[$\Dnubary$] The frequency of $D$ adjusted for
Doppler shift due to the receiver's velocity in the
direction of the beam center at $\Dtime$.
This is used for all purposes other than RFI detection.

\item[$\Dchirp$] The dechirp value (see above) at which $D$ was found.
\end{description}

Pulses and triplets \change{have an additional attribute}:
\begin{description}
\item[$\Dperiod$] The time between consecutive crests.
\end{description}
Autocorr\change{elations have an additional attribute}:
\begin{description}
\item[$\Ddelay$] The delay at which the correlation occurred.
\end{description}

\subsection{Position and frequency uncertainty
\label{subsection:uncertainty}}

Detection positions are uncertain.
Suppose, for example, that an ET signal emanates from a particular sky position $P$.
The signal could result in a detection $D$\change{ from a particular beam}.
The position of this detection, $\Dpos$, is the center of the beam at time $\Dtime$.
Since the beam's sensitivity profile has nonzero width,
$P$ could differ from $\Dpos$.
In fact, the beam has nonzero sensitivity over the entire sky,
so a sufficiently powerful signal could be detected
no matter where the telescope is pointing.
However, ET signals are presumed to have low received power,
so we assume that $P$ is within the beam's half-power disk.

Thus, we define a {\em position uncertainty}, $\Upos = \AObeamwidth$.
If detections lie in a disk of radius $\Upos$ centered at $P$,
it is plausible that they result from an ET signal emanating from $P$.
Detections outside this disk are unlikely to result from such a signal.

Equivalently, if $S$ is a set of detections
and the maximum angle between their positions is at most $2\Upos$,
it is plausible that the detections in $S$ originate from a single ET signal.

Similarly, the frequency of detections is uncertain.
Suppose an ET signal is barycentric:
that is, its transmission frequency is corrected for the Doppler drift due to the transmitter accelerations,
or the transmitter is unaccelerated and no correction is needed.
This signal could result in detections $D$ with barycentric frequencies
$\Dnubary$ that are within an interval of nonzero width.
The deviation can have several sources.

First, the correction for receiver Doppler drift is based on the
direction of the beam center,
but the sky position of the signal may differ from this.
At an angular distance $\Upos$, this results in a maximum error of about 40 Hz.

Second, assuming that the signal is being transmitted in
a beam of nonzero width,
there could be an analogous error in its Doppler drift correction.
It is difficult to estimate the magnitude of such an error without knowing the
distance to and design of the transmitter.
It is likely, however, that deliberate
transmission towards our Solar System would be directed with much better precision
than an Arecibo beam width,
so we estimate this uncertainty to be no more than a few Hz.

Third, recall that the SETI@home front end does coherent
integration with a discrete set of drift rates.
The step size is chosen so that the frequency between successive steps changes by
one-half the bin width over the duration of a bin.
If the actual drift of
the signal due to receiver acceleration lies halfway between steps,
this can result in a frequency difference.
The maximum difference depends on \change{the }DFT length.
For DFT lengths of 128 or more (which include 99.99\% of spikes),
it is 76 Hz.

Summing these uncertainties in quadrature gives an approximate maximum
frequency uncertainty, $\Ufreq = 125\,\Hz$.

If a set of detections has barycentric frequencies lying
in an interval $[\nu - \Ufreq, \nu + \Ufreq]$,
it is plausible that \change{the detections} result from an ET signal
with barycentric frequency $\nu$.
Detections with $\Dnubary$ outside this interval
are unlikely to result from such a signal.

The quantities $\Upos$ and $\Ufreq$ play a key role in the identification of multiplets.
We only consider sets of detections
whose positions and frequencies vary by
at most twice the corresponding uncertainty factor
(see \S\change{\ref{section:multiplets}}).

\section{Observations and sky coverage
\label{section:coverage}}

\subsection{Observation modes\label{subsection:modes}}

SETI@home observed {\em commensally}\change{; that is,}
it did not control the pointing of the telescope during its observations.
\change{P}ointing was determined by other uses of the ALFA receiver:
searching for pulsars near the plane of the Galaxy,
mapping the distribution of hydrogen
in all parts of the Galaxy visible from Arecibo,
and searching for extragalactic hydrogen gas
in isolated clouds and in nearby galaxies. 

The pointing of the telescope may be roughly divided into four \change{modes.}
\begin{description}
\item[Tracking] The telescope tracks a fixed point in the sky.
This mode is typically used for pulsar surveys,
with dwell \change{times ranging} from 30 seconds to tens of minutes \citep{cordes06}.
\item[Drift scan] The telescope is not moving
and the beams move across
the sky at the sidereal rate, $\sim15\arcsec\,\persec$.
Objects in the sky pass through the beam
of one of the ALFA receivers
(0\fdg05) in about 12 seconds.
This mode is typically used for extragalactic hydrogen surveys.

\item[Basket-weave scan] The telescope does a \change{zigzag} scan of the sky,
in a path designed to cover a large area of sky.
The crossing points of this path can be used to determine
the relative gain of the receivers,
\change{since} each is measuring the same astronomical signal at that point. 
The angular velocity is typically near $90\arcsec\,\persec$.

\item[Slewing] The telescope moves rapidly between
distant points in the sky.
The angular velocity can be as low as 0.04\degrees\,\persec\
for pure zenith angle motions and as high as 0.4\degrees\,\persec\
for motions involving the azimuth axis.
\end{description}
See \citet{peek11} for details of the drift scan and basket-weave scan
techniques.

SETI@home's observations alternated unpredictably between these modes. 
Our algorithms that involve pointing
(such as RFI removal and candidate scoring)
were required to work for the full range of pointing trajectories.  \change{The maximum continuous observation at Arecibo was 0.1 day.  Any observations more widely separated than this value were at least 0.9 days apart.  This 0.1 day value was used as a limiting case for some algorithms.}

\subsection{Observation intervals
\label{subsection:observation}}

\change{During} the \alfanyears \change{of }observing,
SETI@home recorded about 400 days of data,
or about 9\% of the total time.
The ALFA receiver was in use only part of the time,
and the SETI@home data recorder was sometimes offline.

Scoring multiplets (see \S\ref{subsection:scoring})
requires knowing the periods during which
we observed each sky position.
For example, \change{suppose that} a multiplet consists of detections from a one-minute span.
If we observed that sky position for only that minute,
the multiplet should score higher than if we observed it for an hour
and found no similar detections in the other 59 minutes.
\change{Calculating} these {\em observation intervals} \change{requires} some approximations.

First, we quantize sky position
using the HEALPix system \citep{Gorski05},
an equal-area pixelization of the sky.
We use the resolution parameter $\rm{N_{SIDE}}=2^{11}$
for which there are $12\times 2^{22}\sim 50{\rm M}$ pixels,
resulting in an average pixel scale of 0\fdg0286,
about half of the ALFA half-power beamwidth of 0\fdg05.
\change{Approximately} 15M of the 50M pixels are visible using the ALFA receiver.

Second, we say that a beam observed a pixel
at a particular time if the beam\change{'s} center
is within $\AObeamwidth$ of the pixel's center at that time.

Third, the rate \change{at which we sample} sky position is only 1 Hz
(see \S\ref{subsection:data_recording}).
During rapid \change{movement,} the sampled pointings can skip over pixels.
To fill in these gaps,
we assume \change{that the} beam position is linear in RA and Dec between samples.

For a pixel $P$ we let $I(P)$ denote the set of time intervals
during which at least one beam observed $P$;
these are the times during which a signal emanating from $P$ could plausibly
be detected in our data.
An algorithm for calculating $I(P)$,
given the above approximations,
is given in Appendix~\ref{appendix:obs}.

\subsection{Sky coverage as a function of frequency resolution}

The narrower the bandwidth of a signal,
the longer the continuous observation needed to detect it
with maximum sensitivity.
Each DFT length $\fftlen$ has an associated duration $\fftlendur$
and frequency resolution $\fftlenres$.
For some pixels $P$ and DFT lengths $\fftlen$,
our longest observation of $P$ is shorter than $\fftlendur$.
Any DFT bin of length $\fftlen$ hence covers a band of sky
that goes outside of $P$,
limiting our sensitivity to signals narrower than $\fftlenres$.
Thus, the effective sky coverage of SETI@home varies with signal bandwidth.

It is possible that a spike with DFT length $\fftlen$
is found during an observation shorter than $\fftlendur$.
In that case, the spike's sky position is a \change{stripe} that is longer than a beam width.
Such spikes are typically RFI.

DFT bins for a \change{DFT length $\fftlen$ begin at regularly spaced times
separated by $\fftlendur$.
Multiplets comprise a}t least two time-disjoint detections
(see \S\ref{section:multiplets}).
\change{Thus, a pixel $P$ can contain} a multiplet composed of spikes of DFT length
$\fftlen$ \change{only if}
our observations of $P$ include at least two intervals
of \change{duration} $\fftlendur$ \change{or more}.

An observation of duration at least $3\fftlendur$ always contains at least two
DFT intervals in their entirety;
an observation of duration at least $2\fftlendur$
always contains at least one DFT \change{interval} and may contain two;
an observation of duration at least $\fftlendur$ \change{might contain only} one.

We say that a pixel $P$ has been ``strongly observed at DFT length $\fftlen$"
if its observation intervals always contain \change{at least} two DFT intervals.
This is the case if $P$ has
an observation \change{with} duration at least $3\fftlendur$, or
two observations \change{with} duration at least $2\fftlendur$. 

We say that $P$ has been ``weakly observed at DFT length $\fftlen$"
if its observations may contain two DFT intervals.
This is the case if $P$ has
an observation \change{with} duration at least $2\fftlendur$, or
two observations \change{with} duration at least $\fftlendur$. 

The fractions of pixels strongly and weakly observed
at the various DFT lengths are shown in
Table \ref{table:coverage_bw}.
For long DFT lengths
(where SETI@home \change{is the} most sensitive; see \S\ref{subsection:sensitivity}),
we have sufficient observations of only a small part of the sky.

\begin{table}[tbp]
\begin{center}
\begin{tabular}{r c c}
DFT length & \% of pixels strongly observed & \% of pixels weakly observed \\
\hline
8 & 100 & 100 \\
16 & 100 & 100 \\
32 & 100 & 100 \\
64 & 100 & 100 \\
128 & 100 & 100 \\
256 & 100 & 100 \\
512 & 99.73 & 99.75 \\
1024 & 97.42 & 98.98 \\
2048 & 93.51 & 96.89 \\
4096 & 78.29 & 91.49\\
8192 & 50.67 & 71.00 \\
16384 & 42.56 & 48.44 \\
32768 & 21.43 & 37.98 \\
65536 & 2.41 & 11.61 \\
131072 & 2.24 & 2.37 \\
\end{tabular}
\caption{Sky coverage as a function of DFT length\label{table:coverage_bw}}
\end{center}
\end{table}

\section{Candidate birdies
\label{section:birdies}}

Signal analysis systems can be tested by injecting artificial signals
and verifying that these signals are reflected correctly
in the output of the system.
This technique is often called ``signal injection and recovery."   
\change{For example, we tested t}he SETI@home front end \change{using} hardware-based signal injection;
this and other validations are described in \cite{instrument_paper}.

Although hardware signal injection is useful for basic functional testing,
it is difficult and expensive for hardware to
generate complex and varied signals.
The signal generation hardware \change{used to test the front end}
can only inject simple signals,
such as sinusoids and noise.   

To test the SETI@home back end, we developed a
sophisticated software signal injector. 
The artificial signals ({\em candidate birdies})
simulate persistent cosmic transmissions with fixed sky positions,
and with a range of parameters: power, bandwidth, planetary motion,
and possible correction for transmitter Doppler shift.
For each \change{birdie,} we generate a set of {\em birdie detections},
mimicking as closely as possible what the
SETI@home front end would produce.
We add these to the detection database at the
start of the pipeline, before RFI removal.

The birdie mechanism serves several purposes,
both in the development of the SETI@home back end
and in its \change{results.}

\begin{description}
\item[RFI filtering] RFI filters can potentially remove ET signals;
birdies let us estimate the extent of this.
Birdie detections do not contain RFI,
but some of them are inevitably removed by RFI \change{filtering,}
for example\change{, detections} that lie in RFI frequency zones.
The fraction of birdie detections that are removed
indicates the likelihood that
our RFI filters would remove a target signal.
As we develop RFI filters, we monitor this fraction
and keep it below about 10\%.
    
\item[Multiplet finding] Birdies help us develop effective algorithms for finding multiplets.
If these algorithms fail to find multiplets for a birdie,
or omit some of its detections,
we can study these cases and improve the algorithms.
    
\item[Score functions] We used birdies to develop our multiplet \change{scoring functions}
by trying to find functions that rank birdie multiplets higher than other multiplets.
    
\item[System sensitivity] By generating birdies with a range of parameters
and seeing which of them are uncovered
(i.e. produce highly ranked multiplets),
we can estimate the sensitivity of
the back-end system as a whole.
    
\end{description}

In this work, we created birdies that model \change{continuous narrowband} signals and generated only spikes for these birdies.
The approach could be extended to model pulsed signals
and generate detections of other types.

\subsection{Birdie parameters}

Each birdie $B$ represents a signal with several parameters:
\begin{description}
\item[$\Bpos$] The signal's sky position (RA, dec).

\item[$\Bfreq$] The center frequency of the signal.

\item[$\Bbandwidth$] The bandwidth of the signal.

\change{\item[$B_{is\_bary}$]{whether the signal frequency is adjusted
(by the sender) to cancel Doppler shift due to acceleration of the transmitter's reference frame.}}

\item[$\BSNR$] The power of the signal as it arrives at the receiver,
in units of signal-to-noise ratio.

\item[$\Bflux$] The power of the signal as it arrives at the receiver,
in \change{\wattpermsqr.}
\end{description}

$F(B)$ is related to $\BSNR$ by

\begin{equation}
\Bflux = \BSNR \Bbandwidth {{\left(T_{\rm sys} + T_{\rm sky}\left(\alpha_{\rm B},\delta_{\rm B}\right) \right)} \over {(\epsilon \Gamma_{\rm B}  \left<M_{\rm bin}\right> \left<M_{\rm beam}\right>)}}
\end{equation}
 where \change{the terms include}
\begin{description}
\item[$T_{sys}$]{the Arecibo/ALFA system temperature (27\,K).}
\item[$T_{sky}\left(\alpha_{\rm B},\delta_{\rm B}\right)$]{the galactic sky brightness at \change{position} $\Bpos$ expressed as a brightness temperature.}
\item[$\Gamma_{\rm B}$]{the \change{geometric gain of Arecibo / ALFA (8 \Kjy).}}

\item[$\epsilon$]{a \change{non-geometric system} loss term dominated by quantization losses ($0.56$, \citeauthor{instrument_paper} \citeyear{instrument_paper}).}

\item[$\left< M_{\rm bin} \right>$]{the median DFT bin response (0.81).}

\item[$\left< M_{\rm beam} \right>$]{the median beam response (0.844).}
\end{description}

A nonbarycentric birdie $B$ has additional parameters: 
\begin{description}
\item[$B_{orbital}$] \change{The} frequency, amplitude and phase of sinusoidal frequency variation
due to the transmitter's orbital motion.

\item[$B_{rotational}$] The parameters \change{of the}
sinusoidal frequency variation due to the \change{rotational motion of the transmitter.}

\end{description}

During the development of the SETI@home back end,
we generated sets of birdies of various sizes
and parameter distributions.
The set of birdies that we used to estimate
sensitivity is described in
\S\ref{subsection:sensitivity}.

\subsection{Generating birdie detections
\label{subsection:birdie_gen}}

Given a birdie -- a virtual ET signal
with parameters as above --
we must approximate the set of detections that
would be produced by the SETI@home front end,
were the signal to exist.
This is a complex task, as we must simulate
the changing Doppler shift of the signal,
the movement and sensitivity of the telescope beams,
the algorithms and parameters of the front-end signal analysis,
the effects of noise, and so on.
We do not model the effects of scintillation.

The algorithm for generating birdie detections
is given in Appendix \ref{appendix:birdies}

The number of detections generated for a birdie depends
on its power
and how closely beam trajectories approach its position.
\change{For some birdies,} no detections \change{are generated.}

\section{RFI removal
\label{section:rfi}}

The Arecibo telescope detects anthropogenic radio frequency interference (RFI).
Sources of RFI \change{include side bands of} TV and radio \change{stations,} aviation radar,
cell phones, and other electronic devices.
RFI can reach a receiver by various paths:
reflection from ionized atmospheric layers,
refraction around the structural components of the telescope,
and direct detection by the receiver and associated electrical systems.

RFI often resembles target signals and is typically more powerful than cosmic sources.
If we did not remove RFI, most of the top-scoring multiplets would be RFI;
it would not be feasible to manually scan these looking for non-RFI multiplets.
Thus, we need to remove as much RFI as possible before finding multiplets.
At the same time, we need to minimize the possibility
of removing target signals.

Many types of RFI are present at Arecibo, each with specific characteristics.
There are a number of radars that produce strong pulsed signals.
The SETI@home front end detects and removes these from the data before
it is analyzed (see \citeauthor{instrument_paper} \citeyear{instrument_paper}).

To remove other types of \change{RFI,} we developed several software filters,
each designed to detect a particular type of RFI.
Most of the filters are based on the fact that the telescope's
sensitivity to RFI does not depend much on pointing:
if RFI is present in a beam at a particular sky position $P$,
it is likely to be present at positions several beamwidths from $P$,
whereas this is not true for cosmic sources.
Thus, if we find groups of detections that are similar
but are separated in position, they may be RFI.
We applied this principle at several times scales:

\begin{description}
\item[Long-term] the entire \alfanyears of observation.
This gave us sensitivity to frequent RFI sources,
even if they are faint.

\item[Medium-term] roughly 10 minutes.
This worked well for intermittent or unstable sources.

\item[Short-term] the timescale of \change{signals detected by the SETI@home client} (1 ms to 13 s).
\end{description}

We developed filters for each of these timescales
and for different \change{types of detection.}
We also used single-detection filters that
flag detections on the basis of their attributes,
with no reference to other detections.

\change{We ran the filters in parallel;
that is, each was run against the entire set of detections,
and the detections flagged by any of the filters were removed.}

\subsection{Long-term RFI filters}

We developed separate filters for \change{continuous detections (spikes and Gaussians),} pulsed detections \change{(pulses and triplets), and} autocorrelations.

\subsubsection{Frequency-zone RFI
\label{subsubsection:zone}}

This type of RFI involves terrestrial signals that
occupy a limited, stable frequency \change{band} and
are present during a significant fraction of the observation period.
Typical sources are TV and radio \change{transmitters}
and leakage from oscillators at the observatory.
This RFI appears as spikes and Gaussians.

We developed an algorithm to identify this type of RFI.
The algorithm divides the overall frequency band into small zones.
For each zone change{$Z$,} we computed the fraction of time $E(Z)$
during which there is a statistical excess of detections in that zone.
We then removed a fraction of the overall frequency band,
consisting of the zones for which $E(Z)$ is greatest.
\change{The algorithm is given in Appendix \ref{appendix:zone}.}

\subsubsection{Period zone RFI}

Periodic detections (pulses and triplets) have a type of RFI
that is analogous to zone RFI, but the zones are in period
rather than frequency.
For example, an aviation radar near Arecibo produces RFI with a $\sim$12 second period;
the injected noise sources used for calibration
typically have on/off cycles of .5, .2, and .1 sec,
and 18 Hz is a common computer interrupt frequency.
So there is a statistical excess of detections
at these periods and their multiples by powers of 2.

The distribution of pulse and triplet periods is not uniform.
They are concentrated at multiples or integer fractions of DFT durations.
The frequency-zone algorithm assumes
a uniform distribution across zones,
so it must be modified to work for period zones.
To do this, we compute the histogram of detection counts
as a function of log period,
smooth it with a \change{30-point} median filter,
and remove period zones whose counts
exceed a probability threshold.

We use a similar algorithm for autocorrelations,
for which the time parameter \change{is the correlation} delay.

\subsection{Medium-term filters}
\subsubsection{Drifting RFI}
We observed in spikes and Gaussians,
\change{narrowband RFI with frequencies that} vary over time.  
An example of this type of RFI is shown in Figure \ref{fig:waterfall}.
This RFI \change{is not} detected by the \change{frequency-}zone algorithm \change{due to} this frequency variation.
We speculate that most of this {\em drifting RFI} is produced by
consumer electronic devices whose oscillator frequencies vary with temperature.

To detect drifting RFI
we adapted an algorithm that had been used in earlier SETI projects \citep{cobb00}.
The basic idea is to construct for each detection $D$,
two {\em fans} of triangles in time/frequency space \change{originating} at $D$,
and extending forward and backward in time; see Figure \ref{fig:drifting}.
The slope of the triangles corresponds to the drift rate of the RFI.
Within each triangle, we count detections \change{whose} sky positions \change{are} sufficiently far from $D$
that they are not likely to have the same source.
Then we look for triangles and pairs of opposed triangles,
with a statistical excess of detections.

Detections often occur in clusters with adjacent or overlapping DFT bins;
these may, for example, be one signal detected at different DFT lengths.
Such clusters can inflate the count of detections in a triangle
and erroneously trigger the algorithm.
So we identify such clusters,
select a {\em master detection} from each cluster,
and use only these detections in the \change{rest} of the algorithm.

\begin{figure}[tb]
\centerline{\includegraphics[scale=.6]{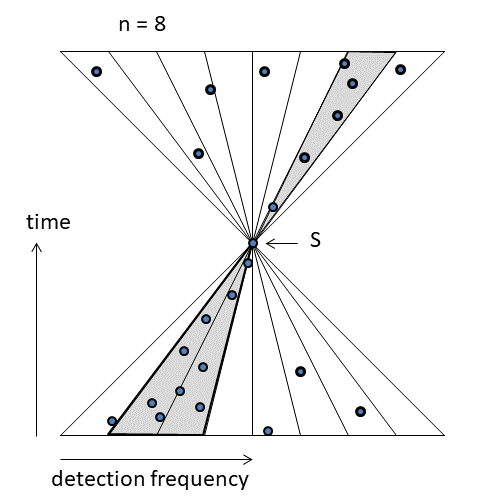}}
\caption{The drifting RFI algorithm constructs fans of triangles in time/frequency space, and looks for triangles containing an excess of detections.
\label{fig:drifting}}
\end{figure}

The algorithm is given in Appendix \ref{appendix:drifting}

Gaussians near spikes flagged as drifting RFI typically are RFI as well,
but often are not flagged as RFI because there
are far fewer Gaussians than spikes.
So \change{we first} compute drifting RFI for \change{spikes,}
save the list of the time/frequency triangles with probabilities
below threshold,
and flag as RFI Gaussians lying in any of these triangles.

\subsubsection{Medium-term pulse and triplet RFI}

The pulsed detection types (pulses and triplets) also have
RFI that occurs on timescales of the order of 10 minutes:
groups of detections, similar in period,
spread out in sky position.

We use the following filter to identify this type of RFI.
It \change{operates on} windows of at most 10 minutes duration,
identifies the detection $D$ at the midpoint of this interval,
and counts the detections that are far from $D$ in sky position
in both the positive and negative time directions.
For typical observation modes,
these two sets are not only far from $D$, but are also far from each other.
If, in a given frequency bin, both sets have a statistical excess of detections,
then they are flagged as RFI.

The algorithm is given in Appendix \ref{appendix:pulse_rfi}.

\subsection{Short-term RFI filter\label{subsection:multibeam}}

If two detections are close in time
and have similar properties (frequency or period, depending on type)
but significantly different positions,
it is likely that they are both RFI.
This motivates the following filter.
\change{Typically,} the two detections are from different beams,
so we call it the {\em multi-beam filter};
however, the detections may be
from the same beam if the telescope is slewing quickly.

The algorithm is given in \change{Appendix} \ref{appendix:mb}

\subsection{\change{Signal property} filters}
Two additional RFI filters consider \change{the properties of each signal rather than the properties of a group of signals}:

\subsubsection{Doppler drift rate consistency}
Because RFI \change{is terrestrial in} origin, it is detected at drift rates close to zero.
Hence we flag as RFI all spikes and Gaussians $D$ for which
\begin{equation}
\left| \Dchirp \right| \lt 0.086\ \hzpersec 
\end{equation}
This is done only for detections with $\Dfftlen \gt 32 {\rm ki}$.
At DFT lengths of $32{\rm ki}$ and less, the frequency bin width
is such that the terrestrial drift rate is indistinguishable from zero.

\subsubsection{Duration consistency}

Each detection $D$ has an associated duration $\Ddur$,
as described in \S\ref{subsection:multibeam}.
In the case of \change{spikes,} this is the duration of its DFT bin.
During this time, the beam $\Dbeam$
is moving \change{at} an angular velocity $V$.
Let $C(V)$ denote the corresponding beam crossing time,
\change{that is,} the time it takes for the half-power beam to cross a sky point.

If $T < C(V)$, then in terms of sky position $D$ is spread
beyond a single half-power beam;
in a sense it is too long for the observation in which it occurs.

For example, \change{suppose that} a \maxtimebin spike $D$
(the longest spike, hence the narrowest bandwidth)
occurs at a \change{moment} when the beam-crossing time is 1 second.
Then it is unlikely that the source of $D$ is a cosmic signal;
it is more likely RFI or noise.
And if $D$ is from a cosmic signal,
that signal will probably be detected with greater
power at a shorter DFT length.

We identify such spikes using an approximate algorithm
based on pixel observation intervals.
Given a spike $D$, let $P$ be the pixel containing $\Dpos$,
and let $b = \Dbeam$.
Recall (\S\ref{subsection:observation})
that $I(P, b)$ is the set of time intervals during
which beam $b$ observed pixel $P$.
Let $I_D$ denote the time interval of duration $\Ddur$
centered at $\Dtime$.
If
\begin{equation}
|I(P, b)\cap I_D| < |I_D|/2
\end{equation}
(i.e. if at least half of $I_D$ lies outside the observation
intervals of its pixel) then flag $D$ as RFI.

About 56\% of spikes were flagged by this filter.
The search for other detection types is
already limited by observation duration,
so no filter is needed.

\subsection{Developing RFI filters\label{subsection:tools}}

We do not know the characteristics of all types of RFI in advance,
so the development of RFI filters,
and the selection of their parameters,
has been iterative and heuristic.
The development cycle is as follows:

\begin{enumerate}
\item Run the current set of filters.
\item From the detections that remain,
find the top-scoring multiplets. (see \S\ref{section:multiplets})
\item Examine these multiplets and the waterfall plots (see below)
of their component detections, looking for RFI.
\item For RFI that should have been removed by an existing filter,
  modify the filter's algorithm or its parameters appropriately.
\item For new types of RFI, study their characteristics and add a new filter.
\item Examine birdie spikes that were flagged as RFI;
in cases where they do not resemble RFI,
\change{modify the} filters to not flag them.
\end{enumerate}

To support this \change{process,} we developed a set of web-based tools.
Using these \change{tools,} one can view lists of top-scoring multiplets
of different types.
For each \change{multiplet,} one can view a list
of the component detections.
For each \change{detection,} one can view a graphical
{\em waterfall plot} showing
the nearby detections in time/frequency space.
RFI is generally visually apparent in these plots.

\begin{figure}[tbp]
    \centering
    \includegraphics[scale=.4]{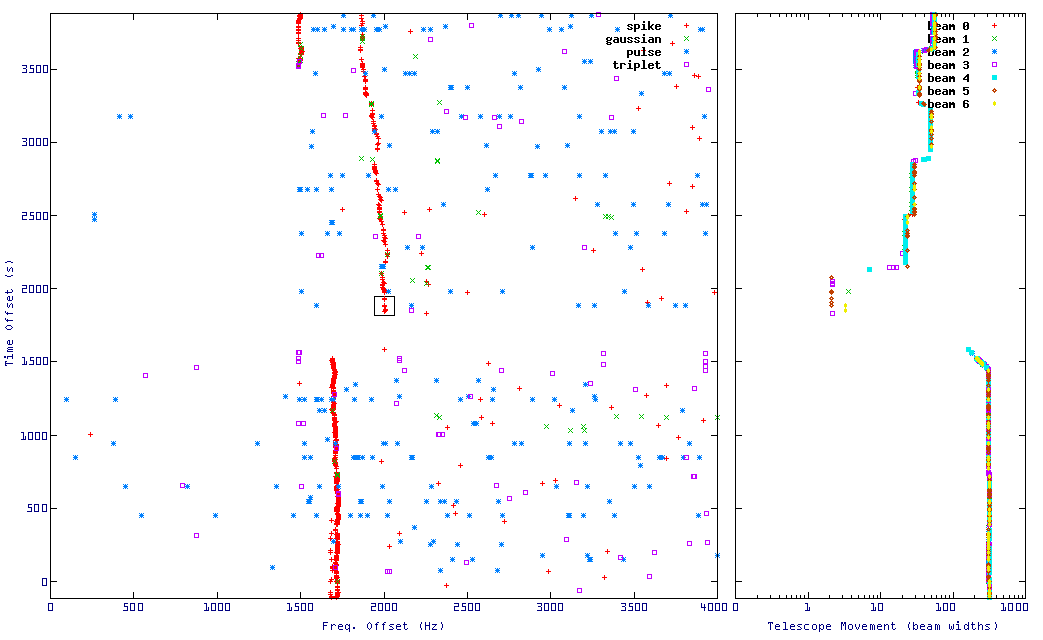}
    \caption{A waterfall plot showing drifting RFI.
    The left panel shows detections in time/frequency space,
    and the right panel shows the distance in sky position
    from the center detection to each of the seven beams. \change{The black box in the left panel is centered on the center detection.}
    \label{fig:waterfall}}
\end{figure}

An example of a waterfall plot is shown in Figure \ref{fig:waterfall}.
The left panel shows the detections, of all types.
The right panel shows the
angular distance (in \change{beamwidths}, on a log scale) of each of
the 7 beams from the sky position of the detection,
as a function of time.
This is useful \change{for} identifying RFI,
since groups of detections close in frequency and time, but at
different sky positions, are generally RFI.

The waterfall plot web interface lets users move up or down, or zoom
in or out, in either dimension.
It lets them view the detections with or without RFI removal.
It also lets them specify parameter
values for the different RFI filters,
and re-run the filters.

The interface also lets users bookmark detections of interest
so that they can be \change{reexamined} later.
This is useful for checking that algorithmic or parameter
changes made for a particular case do not fail in other cases.

\subsection{Evaluating RFI removal}

The fractions of detections of each type
flagged by each RFI filter, and in total,
are shown in Table \ref{tab:fracrfi}.

\begin{table}[tbp]
\caption{Fraction of signals flagged as RFI.\label{tab:fracrfi}}
\begin{center}
\begin{tabular}{c c c c c c c c} 
\hline
Signal type	& Zone & Multibeam & Low drift & Drifting & In spike drifting & All \\
\hline\hline    
all & 7.19\% & 4.46\% & 4.73\% & 5.47\% &  0.06\% & 10.83\% \\
spike & 13.31\% & 6.12\% & 8.44\% & 11.82\% & 0.00\% & 17.49\% \\
Gaussian & 1.69\% & 0.79\% & 0.00\% & 0.24\% & 2.38\% & 4.15\% \\
pulse & 2.33\% & 6.61\% & 0.00\% & 2.15\% & 0.00\% & 7.36\% \\
triplet & 0.17\% & 1.01\% & 0.00\% & 0.01\% & 0.00\% & 1.14\% \\
autocorrelation & 10.67\% & 0.00\% & 13.17\% & 0.00\% & 0.00\% & 14.49\% \\
birdie spikes & 8.78\% & 0.10\% & 0.56\% & 2.90\% & 0.00\% & 11.46\% \\
\hline
\end{tabular}
\end{center}
\end{table}

In evaluating the RFI removal \change{system,} there are two main criteria.
First, it must remove enough RFI so that
a significant fraction of top-ranking multiplets
are not obvious RFI;
otherwise, finding the non-RFI multiplets manually would
take too much time.
Our system \change{has, in fact,} done this; see \S\ref{section:results}.

We can also confirm this by looking at statistical measures.
For example, the distribution of spike frequencies
outside of a narrow band around the hydrogen line
would be uniform in the absence of RFI.
The zone algorithm (\S\ref{subsubsection:zone}) is
designed to enforce this and \change{works} as intended.
Similarly, the distribution of spike power in noise
should be a negative exponential,
while RFI skews this distribution towards higher power.
Indeed, the output of our RFI removal system
has about the right distribution of power.

The second criterion is that the RFI removal system
should not remove target signals.
We used birdies (\S\ref{section:birdies}) -- surrogates for target signals -- to study this.
RFI removal inevitably flags some birdie detections.
For example, because birdie frequencies are random,
some of their detections are in RFI frequency zones.
However, as shown in Table \ref{tab:fracrfi},
the fraction of birdie spikes flagged as RFI (11.36\%) is significantly
less than the fraction of flagged non-birdie spikes (17.49\%),
and as shown in \S\ref{subsection:sensitivity},
sufficient birdie spikes are not flagged
that multiplets are found for most birdies.

\section{Target signal candidates
\label{section:multiplets}}

After RFI removal, the remaining {\em clean} detections consist primarily of
noise from the receiver and astrophysical sources,
birdie spikes,
and possibly target signals (see \S\ref{section:target}).
The next analysis step is to look
for sets of \change{these }detections that could plausibly
be artifacts of a target signal.
This identifies sets for which
a) the detections' frequencies and drift rates are compatible
with one of the target signal \change{types,}
and b) the detections' sky positions are close enough
for them to plausibly have the same source.

We call these sets {\em multiplets}.
The SETI@home ALFA observations span \alfanyears,
and most sky locations have been observed multiple times.
The detections in a multiplet may come
from many observations over a long period of time.

Multiplets are assigned scores designed to reflect the likelihood that
they are the result of a target signal transmission
rather than noise (see \ref{subsection:scoring}).
The multiplet-finding algorithm \change{searches for}
multiplets with high scores.

\subsection{Barycentric and nonbarycentric multiplets
\label{subsection:multiplet_types}}

Recall from \S\ref{section:target} that target signals
may or may not be corrected for transmitter Doppler shift;
if a transmission is corrected in this way, we call it barycentric.

A barycentric signal will arrive at a nearly constant frequency.
However, due to factors described in \S\ref{subsection:uncertainty},
the frequencies $\Dnubary$ of the detections resulting from the signal
can differ from this frequency by up to $\Ufreq$.
So, multiplets arising from barycentric signals
will have detections for which $\Dnubary$
varies by at most $\mdfbary = 2\Ufreq$ = 250 Hz.
We call these {\em barycentric multiplets}.

The frequency of nonbarycentric signals is Doppler shifted
due to the sender's orbital and rotational planetary motion.
For the range of planetary parameters that we consider,
the maximum range of this shift is about \maxsendershift.
The shift resulting from orbital motion varies slowly but
covers a wide range;
the shift resulting from rotational motion varies faster over a smaller range.
See Figure \ref{fig:nonbary}.

\begin{figure}[tbp]
\centerline{\includegraphics[width=12cm]{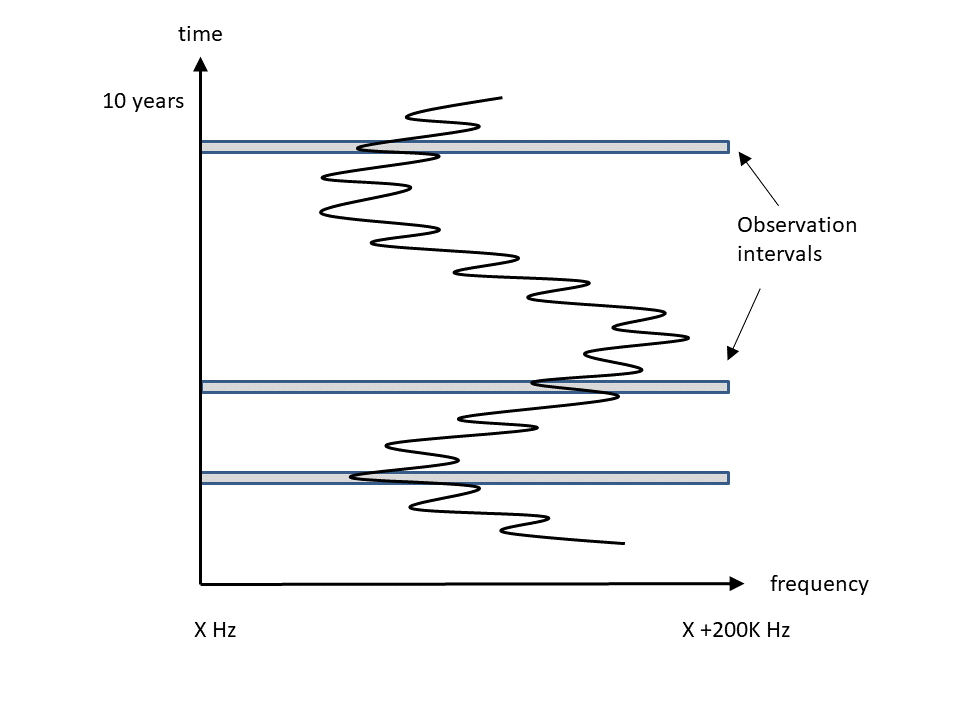}}
\caption{Nonbarycentric multiplets may include detections
over a frequency range of up to \maxsendershift.
\label{fig:nonbary}}
\end{figure}

Thus, the maximum frequency range of nonbarycentric multiplets is $\mdfnonbary = \maxsendershift$.

For the planetary parameters we consider,
variations of this magnitude occur over long timescales
(months or years).
On short timescales \change{(minutes),} sender shifts are small;
detection frequencies can vary
within the uncertainty range of $2\Ufreq$, or 250 Hz.

The techniques for finding and scoring multiplets of
these two classes are somewhat different.
Thus, our algorithm for finding signal candidates
has two variants:
one that looks for barycentric multiplets
in frequency windows of $\mdfbary$,
and one that looks for nonbarycentric multiplets
in frequency windows of $\mdfnonbary$.

\subsection{Multiplet categories}

A \change{continuous narrowband} signal could produce Gaussians
when the telescope moves slowly across its source \change{position,}
and spikes at other times.
To maximize sensitivity,
we form multiplets from the combined set of spikes and Gaussians.
Similarly, a pulsed transmission could produce both pulses and triplets,
so we form multiplets from the combined set of pulses and triplets.
We form autocorrelation multiplets separately.

Thus, multiplets can be from any of three detection sets
(spike+Gaussian, pulse+triplet, and autocorrelation)
and can be barycentric or nonbarycentric.
Each of the six combinations is called a {\em multiplet category}.
The categories differ in many respects;
for example, their score distributions differ.
We generate a separate list of top-scoring multiplets
for each category.

\subsection{Sky position of multiplets
\label{subsection:disk}}

A target signal is assumed to have a fixed sky position.
However, the detections resulting from the signal
will generally have different sky positions due to position uncertainty (see \S\ref{subsection:uncertainty}).

Our multiplet-finding algorithm makes two approximations involving sky position.
First, when we form multiplets, we require that
the sky positions of the detections lie in a disk
of radius $\AObeamwidth$.
This ensures that all detections in the disk
are within the uncertainty radius $\Upos$ of its center
and therefore could be from the same source.

Second, we consider only disks centered at the center of HEALPix pixels.
For a pixel P, ${\rm disk}(P)$ denotes the set of sky positions
centered on $P$ and with radius $\AObeamwidth$.
\change{These} disks overlap (see Figure \ref{fig:disk}).
Note that the set of detections resulting
from a target signal might not be
contained in a single pixel disk,
in which case we will not necessarily find the best multiplet
for that signal.

\begin{figure}[tbp]
\centerline{\includegraphics[scale=1]{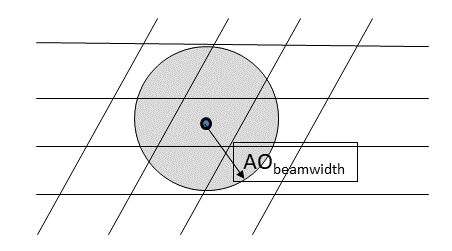}}
\caption{A pixel disk is centered at a pixel
and has radius $\AObeamwidth$.
\label{fig:disk}}
\end{figure}

\subsection{Constraints on multiplets\label{subsection:constraints}}

To ensure that multiplets plausibly result from a target signal,
we require that they satisfy a number of constraints.

\subsubsection{Overall frequency range}

As explained in \S\ref{subsection:multiplet_types},
the detections $D$ forming a multiplet must
have values of \Dnubary that
lie in a range of $\mdfbary$ (250 Hz) for
barycentric multiplets
and $\mdfnonbary$ (308 kHz) for nonbarycentric multiplets.

\subsubsection{Drift rate constraints
\label{subsubsection:chirp_constraints}}

The terrestrial drift rate in the SETI@home band \change{(1418.75--1421.25 \MHz)} at AO ranges from $-0.12$ to $-0.16\,\hzpersec$.
Most of the variation is due to the difference between the observation direction and the rotational acceleration vector.
For barycentric multiplets, we constrained the reported detection drift rate to those consistent with this drift range, given the drift rate step used at each DFT length.

When determining whether a detection is consistent with having a stable barycentric frequency,
we compare the reported drift rate with the range of allowed drift rates for a signal detected at that DFT bandwidth.  
For the 128ki DFT length, the allowed range is -0.19 \hzpersec to -0.09 \hzpersec.
For DFT lengths shorter than 16ki,
any nonzero drift rate disqualifies the detection.

\subsubsection{Period and delay consistency
\label{subsubsection:period_consistency}}

For pulse/triplet multiplets, we require that
the periods of the detections be about the same,
and similarly for the delays in autocorrelation multiplets.
For pulse/triplet multiplets,
two detections are considered {\em consistent}
if their periods differ by at most
twice the \change{maximum} of the two DFT \change{durations.}

For barycentric autocorrelation multiplets,
two detections are considered consistent if
their delays differ by at most
$\mdpbary$ = 0.01 s.

For nonbarycentric multiplets (for which the delays
may be Doppler-shifted), the delays must differ by at most
$\mdpnonbary$ = 0.1 s.

\subsubsection{Local drift rate/frequency consistency for nonbarycentric multiplets
\label{subsubsection:local_nonbary}}

\change{Given our assumptions about target signals (\S\ref{section:target}),}
planetary motion \change{on timescales of hours or less}
causes only small changes in drift rate.
Therefore, if two detections are close in time and have \change{very}
different drift rates, they are not from the same source.
In addition, if a detection $D$ has a drift rate of 1\,\hzpersec,
we would expect a detection $D^\prime$ 10 seconds later
to have a frequency approximately 10 Hz higher.
The more it differs from this,
the less likely that $D$ and $D^\prime$ \change{have} the same source.

To express \change{this,} we impose a constraint
with the following parameters:

\begin{description}
\item[$\mdtlocal$] The time period over which we enforce consistency: 0.1 day.
Over this duration, drift rate is fairly constant
for our range of target signals.
\item[$\mdclocal$] The maximum allowed variation in drift rate
over periods of duration $\mdtlocal$: 10\,\hzpersec.
\item[$\mdflocal$] The maximum  
variation in drift-adjusted frequency during that period:
$2\Ufreq$, or 250 Hz.
\end{description}

\change{A drift rate change of 10\,\hzpersec corresponds to a change in the
line-of-sight acceleration of $2.2 \unit{m s^{-2}}$.
While this could potentially occur in a long observation of a short period satellite, because the vast majority of our observations are significantly less than one minute it is unlikely to occur in practice.}

\subsubsection{Global drift rate/frequency consistency for nonbarycentric multiplets\label{subsubsection:global_nonbary}}

We require that the change in Doppler drift rate and frequency
over longer time periods be consistent with planetary motion
of the range we are considering.
We approach this problem by asking:
If (given our assumptions on planetary motion)
a signal has drift-rate and frequency $(C_1, F_1)$ and time $t_1$,
is it possible that it has drift-rate/frequency $(C_2, F_2)$ at a later time $t_2$?
Based on an analysis of a large number of nonbarycentric birdies,
we empirically found that the inequality
\begin{equation}
|C_1 - C_2|\ |F_1 - F_2| < 0.75 \tten{7}\sqrt{t_2 - t_1} 
\end{equation}
provides a fairly tight bound for our range of planetary and stellar parameters.

We use this as our global consistency constraint.
This is necessary but not sufficient for the multiplet
to match an actual orbital/rotational system.
It does not ensure that the detection frequency and drift rate fit
plausible sum-of-sinusoids functions
that would be expected of target signals.

\subsubsection{Frequency-variation consistency for barycentric multiplets
\label{subsubsection:freqvar}}

The detections comprising a barycentric multiplet
should vary little in barycentric frequency,
but their topocentric frequencies can (and should) vary
because of receiver acceleration.

At one point, we noticed a number of high-scoring
barycentric multiplets for which the opposite was true:
There was more variation in barycentric frequency than
in topocentric frequency.
This is evidence that these groups were due to a frequency-stable terrestrial source.
Often these multiplets resembled weak drifting RFI but did not have
sufficient detections to trigger the drifting RFI algorithm.

To disallow such multiplets, we added the following constraint
for spike/Gaussian barycentric multiplets.
Such a multiplet may not contain a group of detections that
are within a 0.1 day time interval (about the maximum amount of time that a celestial source could be observed continuously from Arecibo),
and for which
\begin{equation}
B > 2 \max(R, T)
\end{equation}
where $B$ is the RMS variation in barycentric frequency,
$T$ is the RMS variation in topocentric frequency,
and $R$ is the coarsest frequency resolution
of the detections in the group.

\subsubsection{Time disjointness of detections
\label{subsubsection:time_disjoint}}

Recall (\S\ref{subsection:detections})
that each detection occupies a particular time interval.
A cosmic signal may produce detections for
which these intervals overlap: perhaps spikes in frequency-adjacent DFT bins or of different DFT lengths.
These detections are redundant,
and including \change{all of them} in a multiplet would inflate its score.
So we require that a multiplet's detections not overlap in time;
in other words, that the intervals determined by
$\Dtime$ and $\Ddur$ not overlap.

\subsubsection{Frequency disjointness of multiplets
\label{subsection:disjoint}}
Our initial implementation of multiplet finding
produced a large number of multiplets differing by only one or two detections.
This is undesirable; it fills the high-score lists with
copies of essentially the same multiplet.
Ideally, there should be one multiplet for
a given target signal candidate.
So, we require that multiplets of a given category be disjoint.
We enforce this with a stronger constraint:
If $M_1$ and $M_2$ are multiplets of the same category,
the ranges of the frequencies of
their detections \change{(that is,} $\Dnubary$) must be disjoint.

\subsubsection{Multiple detections}

Finally, all multiplets must include at least two detections.
We originally allowed one-detection {\em singlets},
but these seemed to be entirely RFI.
Only a small fraction of RFI detections leak
through the RFI rejection algorithms,
but these detections are
highly selected when searching for the best singlets.  

Because 97\% of the Arecibo sky is covered two or more times in
our commensal sky survey,
and the median number of observations of a pixel is 13,
our requirement that a signal be detected two or more times
does not significantly lower the probability of detection
of a high duty cycle signal.   

\subsection{Finding nonbarycentric multiplets
\label{subsection:nonbary_multiplet}}

We now describe how we find multiplets:
That is, given a pixel disk of detections of a given category
(spike/Gaussian, pulse/triplet, or autocorrelation)
how we find subsets of these detections that satisfy the above constraints.
We do this in a way that tries to maximize the
scores of the resulting multiplets.
In practice, this means trying to find multiplets
that have as many detections as possible
and whose detections have the highest scores as possible
(although these goals conflict in some cases).

We start with the following subproblem:
Given the set $S$ of detections in a given sky disk
and in frequency band of $\mdfnonbary$
(the maximum width of a nonbarycentric multiplet),
what is the subset $T \subseteq S$ that

\begin{enumerate}
\item satisfies the consistency constraints described in \S\ref{subsection:constraints}
\item has the highest multiplet score subject to 1).
\end{enumerate}

Finding the highest scoring subset is probably not feasible.
Brute force search -- examining all subsets of $S$ --
uses computing time exponential in the size of $S$,
which can exceed one million.
So we use a heuristic algorithm: starting with $S$,
we prune detections in several stages,
in a way that satisfies the constraints
while maximizing the sum of detection scores.
This \change{increases} the multiplet's {\em power factor},
one of three independent multiplet scores
(see \S\ref{subsection:scoring}).
A more sophisticated algorithm might try to \change{increase} the other factors as well.

\subsubsection{Local drift-rate/frequency pruning
\label{subsubsection:chirp_freq_prune}}

To enforce the local drift-rate/frequency constraint
(see \S\ref{subsubsection:local_nonbary}),
we use an algorithm that,
given a set of detections in a time window $\mdtlocal$,
returns a subset that satisfies the constraint.
It tries to find a subset that will result in a high multiplet score.
This algorithm is given in Appendix \ref{appendix:local_drift}.

The above algorithm produces a locally consistent group of detections
in a short time window.
We then need to assemble multiple such groups
across the full time range in a way
that is globally consistent
(\S\ref{subsubsection:global_nonbary}).
An algorithm for this is given in Appendix \ref{appendix:global_drift}.

\subsubsection{Period and delay pruning}

For pulse/triplet and autocorrelation \change{multiplets,}
we enforce the period/delay consistency constraint
(see \S\ref{subsubsection:period_consistency})
by removing detections.
As with drift-rate/frequency pruning, we want to keep as many detections as possible,
especially high-scoring ones.
We use a similar algorithm:
we sort the detections by period or delay,
then scan this list and find the interval
over which a) the detections have consistent periods or delays
and b) the sum of detection scores is greatest.

\subsubsection{Time overlap pruning
\label{subsubsection:timeoverlap}}

To enforce the \change{time-disjointness} constraint
(\S\ref{subsubsection:time_disjoint}),
we must remove overlapping detections.
Specifically, given a set $S$ of detections,
some of which may overlap in time,
we want to find $S^\prime\subseteq S$ for which no detections overlap in \change{time}
and which maximizes the sum of detection scores.

This is equivalent to the scheduling problem known as
the Weighted Activity Selection Problem:
given a set of {\em activities}, each with a value
and start and end times, find the non-overlapping subset
with the greatest total value.
There is an efficient ($O(Nlog(N)$) algorithm to
solve this problem, using dynamic programming
\change{\citep{Cormen2022}.}

\subsection{Finding barycentric multiplets
\label{subsection:bary_multiplet}}

We have now described the algorithm for finding nonbarycentric
multiplets in a given frequency band,
from the detections in a pixel's disk.
The analogous algorithm for barycentric multiplets
is somewhat \change{simpler} and is given in Appendix \ref{appendix:find_bary_freq}.

\subsection{Finding multiplets in a detection disk}

We have described the algorithms for finding barycentric and nonbarycentric
multiplets in a particular frequency band;
Appendix \ref{appendix:find_bary_disk} describes the algorithm for finding multiplets across all frequencies.

The algorithm for finding nonbarycentric multiplets in a detection disk
is identical, except that the frequency band is $\mdfnonbary$,
the drift rate constraint for nonbarycentric multiplets is used
(\S\ref{subsubsection:chirp_constraints}),
and the algorithm described in \S\ref{subsection:nonbary_multiplet}
is used.

\subsection{Pruning overlapping multiplets across pixels}

For a given pixel, the algorithm in the previous
section generates multiplets
that are disjoint in frequency range
and therefore disjoint in terms of detections
(see \S\ref{subsection:disjoint}).
However, it is possible that
multiplets generated for nearby pixels are not disjoint.
This can fill the top-ranked lists
with several copies of essentially the same multiplet.

A final stage in multiplet-finding identifies
pairs of multiplets in nearby pixels that are of the same category
and have one or more detections in common.
It removes the lower-scoring multiplet from each such pair.

\subsection{Scoring multiplets
\label{subsection:scoring}}

The algorithms described in the previous subsections produce
tens of millions of multiplets.
We extracted about one thousand of these
for manual examination and possible reobservation.
For this purpose, we developed three functions,
or score factors, that take a multiplet $M$ and return
a score -- a measure of a property
that we would expect to find in ET signals and not in noise.
We used these scores to decide which multiplets to examine.

Each factor estimates a probability that $M$ would
be found in noise, assuming Poisson statistics.
These probabilities can be very small,
so we do the computation in log space to avoid loss of numerical precision.
We then negate the log values so that larger values are better.

\subsubsection{Power factor}

Recall from \S\ref{subsection:detections}
that a detection has a probability score $\Dscore$.
The first multiplet score factor, $S_{\rm prob}(M)$, rewards
(i.e. gives high scores to) multiplets
that have detections with high scores,
for example because they have high power.

\Sprob is essentially the median score
of the detections in $M$.
However, the distribution of $\Dscore$ varies
significantly \change{between the} detection types,
because it is defined differently
and the number of detections varies between types.
Millions of spikes have scores higher than
the highest Gaussian score.

To ensure that \Sprob selects
multiplets containing Gaussians,
we normalize scores across detection types.
For each detection \change{type,} we find the
score of the 30 millionth highest detection
and subtract this from the scores of
detections of that type.
\Sprob is then defined as the median
over the detections in $M$
of these normalized scores.
This has the desired effect:
in \change{the} high-ranking multiplets of the mixed types
(spike/Gaussian and pulse/triplet)
both types are well represented. 

\subsubsection{Density factor}

Suppose that for a pixel $P$, ${\rm disk}(P)$
contains $N$ detections of a particular type.
If $A$ is a section of sky contained within ${\rm disk}(P)$,
with area $X {\rm area}(P)$ for some $X < 1$,
then the expected number of detections positioned in $A$
is $N X$.
Similarly, if a frequency band $F$ is contained in
SETI@home's \fullband band,
and its width is $Y \times \fullband$ for some $Y < 1$,
the expected number of detections in $disk(P)$
with frequencies in $F$
is $N Y$,
and the expected number of detections with positions in $A$
and frequencies in $F$ is $N X Y$.

We would expect an ET signal to produce a multiplet $M$
with many detections closely spaced in both
position and frequency.
Suppose a multiplet $M$ is in a pixel disk, ${\rm disk}(P)$,
with $N$ detections,
and the positions of $M^\prime$s detections cover
(in the sense defined below) a
fraction $X(M)$ of the area of ${\rm disk}(P)$,
and the frequencies of $M^\prime$s detections cover
a fraction $Y(M)$ of SETI@home's 2.5 MHz range.

The expected number of detections in these ranges
of positions and time is then
\begin{equation}
E = N X(M) Y(M)
\end{equation}
The probability of finding at least $|M|$ detections
in these ranges is then
\begin{equation}
P(M) = \Gamma(|M|, E)
\end{equation}
where $\Gamma$ is the incomplete Gamma function.

We define the density factor as
\begin{equation}
S_{density}(M) = -\log(P(M))
\end{equation}

The density factor rewards multiplets with more detections
than would be
expected given their range of sky position and frequency.
Thus, it gives high scores to multiplets
that have a large number of detections
and are compact in both position and frequency.

We now define the fractional coverage factors $X(M)$ and $Y(M)$.
Let $A$ be the mean sky position of the detections in $M$,
let $B$ be the standard deviation of the angle between
$A$ and the positions of detections in $M$.
X(M) is then the area of a disk of radius $B$
divided by the area of the sky disk ${\rm disk}(P)$.

To define $Y(M)$,
we first define the {\em frequency deviation}, $D_{dev}$,
of a detection $D$ in a multiplet as
the difference between $\Dnubary$
and the {\em center frequency} of the multiplet at time $\Dtime$.
For barycentric multiplets the center frequency is constant;
it is defined as the mean of the $\Dnubary$ over
all the detections in the multiplet.
For nonbarycentric \change{multiplets,}
the center frequency varies linearly over time within an observation interval;
it is the center of the time/frequency bands
computed during drift-rate/frequency pruning.
(See \S\ref{subsubsection:chirp_freq_prune})
Note that in all cases the frequency deviation
is at most $2\Ufreq$.
$Y(M)$ is then the standard deviation
of $D_{dev}$ divided by 2.5 MHz. 


\subsubsection{Time factor}

A multiplet $M$ consists of detections
\change{within} the disk centered at a pixel $P$.
We know $I(P)$, the set of time intervals during which we observed $P$
(see \S\ref{subsection:observation}).
If $M$ results from a beacon that is always on,
it could contain detections from throughout these intervals.
The {\em time factor} $S_{time}(M)$ attempts to quantify the extent
to which it does.
We do this by estimating the probability that a random set of detections,
with constraints similar to those of multiplets,
would cover at least as much time as the detections in $M$.

We introduce some terms \change{to formalize this}
We call a set of non-overlapping time intervals an {\em interval set}.
If $I$ is an interval set, $D(I)$ denotes its duration,
that is, the sum of the lengths of its component intervals.

Recall that detections have associated time intervals,
centered at $\Dtime$ and of duration $\Ddur$.
The detections in a multiplet $M$ have non-overlapping intervals.
Let $I(M)$ denote the corresponding interval set
and $D(M)$ its duration.

If $P$ is a pixel, let $I(P)$ denote the interval set
consisting of its observations
and $D(P)$ the corresponding duration.
Typically, if $M$ is a multiplet from pixel $P$,
then $I(M) \subseteq I(P)$ and therefore $D(M) < D(P)$.

The detections in a multiplet lie in
a frequency band of width $2\Ufreq$ (250 Hz)
(for barycentric multiplets, this band is fixed;
for nonbarycentric multiplets, it changes over time).
For a multiplet of type $T$ (spike/Gaussian, pulse/triplet, or autocorrelation),
let $F(T)$ denote the average fraction of
time that a 250 Hz band contains detections of type $T$.
We can estimate this as follows:
enumerate \change{the} detections of type $T$ in time order.
We group them into contiguous sets
in which the gap between detections does not exceed 30 seconds.
For each such time interval, divide the detections into 250-Hz bins.
Within each bin compute the amount of the interval that is
covered by the detections.
$F(T)$ is then the average coverage over all bins and time intervals.

We assume that for a given interval set $I$ and 250-Hz band $B$,
the coverage of $I$ by detections in $B$
has a roughly Poisson distribution with mean $F(T) D(I)$.

We can now define the time factor for a multiplet $M$:
\begin{equation}
X = \Gamma(D(M), F(T) D(P))
\end{equation}
\begin{equation}
S_{time}(M) = -log(X)
\end{equation}

where $P$ is the pixel \change{that contains} $M$.
$X$ is the probability that the coverage of $I(P)$ by
a random set of detections of type $T$, within a 250 Hz window,
would exceed the coverage of $M$.

\subsubsection{Evaluating and combining score factors}

We have defined three {\em score factors} \change{designed to measure the multiplet properties that} we expect ET signals to have.
Two questions arise:
\begin{itemize}
\item Do the factors, as we intend, rank ET signals higher than noise?
\item How do we combine the factors into an aggregate score?
\end{itemize}

For the spike/Gaussian \change{categories,} the birdie mechanism
gives us a way to answer the first question:
we can see whether birdie multiplets score higher than non-birdie multiplets.
In fact, this idea was used to develop and refine the score factors.
\change{Given two alternative functions,} we chose the one that
more favors birdie multiplets.

We originally combined the three scoring factors into a single score.
Since each factor is conceptually a probability,
it was natural to multiply them to form a single score.
However, some factors have
a much wider range than others
and they dominated the product.
So we experimented with using different weights for the factors
in an effort to boost the scores of birdie multiplets relative
to non-birdie multiplets.
We \change{tried using both neural networks and} optimization.
Neither \change{approach} was successful\change{}; when we trained with a set of
birdies, the resulting weights worked well for those birdies
but poorly for others.}

\change{Currently,} we use a simpler approach.
First, we linearly scale each factor so that
the 25th percentile and 75th percentile are -1 and 1 respectively.
The score factors have different distributions
for different categories,
so we do this scaling separately for each category.

Second, we consider sums of just one or two of the \change{score} factors,
as well as \change{the sum of} all three.
There are seven such combinations; we call them {\em score variants}.
An ET signal could score high
in one or two factors but not in the others.
For example, a weak but persistent signal could score
high in time and density factors but low in power factor.
A strong but brief signal would do the opposite.

To select multiplets to \change{reobserve,}
we manually examined the top multiplets from each score variant.

\subsection{Evaluating the multiplet-finding and scoring algorithms}

The algorithms for finding and ranking multiplets
described in this section
went through many stages of development and refinement.
For \change{continuous narrowband} signals, this process was guided by birdies.
We evolved the multiplet-finding algorithms
to find most of the non-overlapping detections in a birdie,
and few extraneous detections;
we evolved the scoring functions to rank birdie multiplets
high compared to real multiplets.
This process resulted in algorithms that perform these functions well.
As will be shown in \S\ref{subsection:sensitivity},
they successfully uncover most birdies.

For other detection \change{types,}
it is hard to quantify the performance of the algorithms
because we have no birdies of those types.
In \change{general,} we have proceeded empirically.
We manually examined the top-ranking multiplets.
For each multiplet, we examined the detections
in its range of frequency and/or period,
and \change{made sure that we were} including those we should.
We examined its score factors and determined whether
its high scores were merited.
By these subjective criteria, our algorithms work well.

\section{Implementation and performance
\label{section:implementation}}

The algorithms comprising the SETI@home back end
have been developed empirically in many iterations.
Each iteration required running some or all of the pipeline.
An early version of the back-end pipeline,
called NTPCkr, \citep{Korpela11a,Korpela11c}
took years to run, making algorithm development infeasible.
Starting in 2016 we redesigned the back end,
with the goal of reducing the time for a complete run to about a day.

\subsection{Data storage and I/O}

Much of the back-end processing is data intensive.
The SETI@home front end stores its output in a relational database.
Information is stored in a hierarchy of tables
(see Table \ref{table:table_sizes}).
A row in each table links to a parent row in the next higher table.
Retrieving data from this database is slow.
Traversing links up the table hierarchy
compounds the problem.

\begin{table}[tbp]
\begin{center}
\begin{tabular}{|c|c|c|c|}
\hline
table & each row represents & number of rows \\
\hline\hline
tape & Several hours of data from a beam & 117,466 \\
\hline
workunit group & 107 seconds of data from a tape & 7,691,384  \\
\hline
workunit & One of 256 frequency subbands of a workunit group & 
1,968,994,304 \\
\hline
detection & A detection (spike, Gaussian, etc.) & 12,107,039,965 \\
\hline
\end{tabular}
\end{center}
\caption{Database tables and row count.
\label{table:table_sizes}}
\end{table}

We speed up data access as follows.
First, we dump the relational database to
comma-separated value (CSV) files;
the database is not used further.
We then flatten the table hierarchy.
For example, given a detection, we need the angle range of
its workunit group.
For this purpose, we create an array, indexed by workunit group ID,
of the angle range of that workunit group.
We store this array in a file, accessed using \change{memory mapping.}
Given a detection, we can find the angle range
by using the detection's workunit group ID as an index into this array.
This involves a single memory reference rather than
a chain of database queries.

We need to access detections in time order during RFI removal
and by pixel number during multiplet finding.
We do this in each case by sorting the detection files
on that parameter (using the Unix ``sort" \change{utility),}
then building an index that allows random access based on the parameter.
The RFI detection program makes a single pass through the time-ordered
list of detections of a given type,
minimizing disk I/O.
Multiplet finding makes a single pass through frequency-ordered
detections.

\subsection{Efficient data structures and algorithms}

Many of the algorithms involve processing
$N$ items where $N$ is on the order of a million.
To avoid $O(N^2)$ runtime, the back-end programs use a data structure
called R-trees \citep{Guttman84}.
An R-tree stores a set of geometries such as polygons or points,
and allows the set to be queried
(e.g. to see if a given point is in any of a set of $N$ rectangles)
in $O(log(N))$ time.

For example, while generating birdie detections
(\S\ref{subsection:birdie_gen}),
we store the set of birdie sky positions in an R-tree.
Given a telescope pointing, we can then efficiently identify
the birdies that are close to it.
In RFI removal (\S\ref{section:rfi})
we use R-trees to store drift triangles
and detection uncertainty rectangles.

\subsection{Parallelism}

RFI removal is done by a \change{multithreaded} program;
on a machine with $N$ CPUs, each CPU processes a time range
containing about $1/N$ of the detections.
With 10 billion detections and 56 CPUs,
RFI removal takes about 15 hours.

Multiplet finding and scoring \change{are performed} by
a computing cluster with several thousand nodes, using HTCondor \citep{Thain05}.
The task is divided into jobs of 64 pixels each;
there are about 250,000 jobs.
Each job takes an average of 1120 seconds,
so with 2000 CPUs the task takes about 1.6 days.

\section{Results\label{section:results}}

\change{Once the algorithms were finalized,}
we performed two complete runs of the SETI@home back end.
For the first run, the input was the detections generated
by the SETI@home front end.
The result of this run is a set $M_R$ of {\em real multiplets}.
Table \ref{table:mp_stats} shows some statistics of these multiplets.

\begin{table}[tbp]
\begin{center}
\begin{tabular}{r r l r r}
Multiplet category & \#multiplets & avg. \#detections & avg time span (sec) \\
\hline
spike/gaussian, barycentric & 486092 & 3.51 spikes + 0.07 Gaussians & 110.44 \\
spike/gaussian, nonbarycentric & 2260066 & 2.56 spikes + 0.20 Gaussians & 1,152.81 \\
pulse/triplet, barycentric & 16668902 & 0.11 pulses + 2.00 triplets & 666.19 \\
pulse/triplet, nonbarycentric & 134247 & 0.00 pulses + 2.09 triplets & 890.29 \\
autocorr, barycentric & 287 & 2.13 autocorrs & 53.83 \\
autocorr, nonbarycentric & 880346 & 2.08 autocorrs & 1,103.20 \\

\end{tabular}
\caption{Multiplet statistics.\label{table:mp_stats}}
\end{center}

\end{table}

We manually evaluated the top-ranking multiplets,
skipping those consisting of obvious RFI;
see \S\ref{subsection:mp_select}.
The result of this is a set of sky positions and
frequency ranges to be reobserved.

Next, we generated a number of birdies
\change{that span} our range of \change{continuous} target signals
and generated detections as described in
\S\ref{subsection:birdie_gen}.
We combined these detections with the real \change{detections}
and processed this with the back-end pipeline.
We used the results to estimate the sensitivity
of SETI@home to signals with various characteristics.
To do this, we checked which birdies were uncovered:
i.e., which one produced a multiplet that ranked in the top
100 non-birdie multiplets for some score variant.
For each type of birdie, there was
a power level above which most birdies were uncovered.
This minimum power level, or {\em candidate sensitivity}, depends
on the intrinsic bandwidth of the birdie
and whether it is barycentric.

\subsection{Birdie generation}

For the second run, we created 3000 birdies
with parameters chosen as follows:
\begin{itemize}
\item
\change{Half of} the birdies (1500) are barycentric, half are nonbarycentric.
\item 
$\Bbandwidth$ is chosen randomly in a way
that distributes the values evenly (in log space)
over SETI@home's range of frequency resolution.
For each of the 15 DFT lengths $\fftlen$, there are 100 barycentric
and 100 nonbarycentric birdies with bandwidths
in the range of resolutions closest to $\fftlenres$.
\item
For a birdie with $\Bbandwidth$ closest to DFT length $\fftlen$,
\change{the position $\Bpos$ was} chosen randomly from pixels that
were strongly observed at that DFT length
(see \S\ref{subsection:observation}).
In other words, we \change{placed the} birdies only in areas of the
sky where we had \change{long enough} observations
to detect them with maximum sensitivity.
\item
For nonbarycentric birdies, the planetary
motion parameters were chosen randomly from
ranges consistent with habitable-zone planets
orbiting type F and G stars.
We assume that the signal is always on,
so that if the transmitter is on the surface of a planet,
there is a second transmitter at the antipode.  
\item
$\BSNR$ is between 18 and 33 for barycentric birdies,
and 18 and 50 for nonbarycentric birdies.
Within the 100 birdies in each bandwidth range,
the powers are regularly spaced within these limits.
\item
$\Bfreq$ is chosen randomly from the \fullband band
with a band of $\mdfbary$ Hz (barycentric)
or $\mdfnonbary$ Hz (nonbarycentric) removed from both ends;
this ensures that all detections generated for the birdie
lie in the \fullband band.

\end{itemize}

We generated spikes for the birdies as described
in \S\ref{section:birdies}.  \change{This work predates the publication of work by \citet{li22}, but is generally consistent.  Because \citeauthor{li22} concentrated on exoplanets around less massive stars, the orbital periods of the planets were shorter and the rotational periods typically longer due to tidal resonance.  The equivalent drift rates for 
non-barycentric signals from \citet{li22} 
are well within the SETI@home range, and would not be excluded from being
detected.  However, the
high orbital velocities expand the frequency range of the signals,
making detection of multiplets more difficult within a limited number of observations.  In essence, the number of non-birdie detections included in a multiplet is proportional to the orbital velocity, whereas the number of real signals would be constant.  Hence we limited our birdie generation to orbits around more sunlike stars and acknowledge that uncorrected transmissions from rapidly accelerating systems is difficult.}
These spikes were combined with real spikes prior to RFI removal.
We ran the multiplet-finding algorithm for pixels
containing at least one birdie \change{spike}
and retained only multiplets having at least one birdie spike.
For each birdie $B$, $M(B)$ denotes the set of multiplets in $M_B$
\change{that contain} at least one spike from $B$;
these are collectively called {\em birdie multiplets}.

Each multiplet is scored using the
three score factors described in \S\ref{subsection:scoring},
and these are combined into seven score variants.
For each multiplet $M$ and score variant $V$,
$S(M,V)$ denotes the score of $M$ in variant $V$.

For each birdie multiplet $M$ and each score variant $V$,
$rank(M, V)$ denotes the rank of $S(M, V)$
relative to the scores of the real (non-birdie) multiplets.
For example, $rank(M, V)$ is zero if its $V$ score is greater
than that of all real multiplets.

\subsection{Effectiveness of multiplet score factors and variants
\label{subsection:effectiveness}}

In \S\ref{subsection:scoring}, we defined three
multiplet score factors: $S_{power}$, $S_{density}$, and $S_{time}$.
These were designed to measure properties of
multiplets that are present in target signals more than in noise.
How well do they do this,
and which factors or combination of factors work best?

We studied this for \change{continuous signals}
by examining the ranks of birdie multiplets
and seeing which score variant produced the best ranks.
Specifically:
for a birdie $B$ and score variant $V$, let $rank(B, V)$ denote
the minimum of $rank(M, V)$ over multiplets $M$ containing detections from $B$.
For a score variant $V$, let $F(V)$ be the fraction of birdies
for which $rank(B, V)$ is minimal over score variants;
i.e. the birdies for which $V$ is the best selector.

\begin{table}[tbp]
\begin{center}
\begin{tabular}{c c c c c c c c}
Birdie BW & All & power+time & density+time & time & density+power & power & density \\
\hline
\multicolumn{8}{c}{Barycentric birdies} \\
All & 11.27\% & 4.29\% & 13.46\% & 6.34\% & 22.54\% & 0.05\% & 42.06\% \\
\lt 0.11 Hz & 19.63\% & 15.89\% & 22.90\% & 19.16\% & 7.48\% & 0.00\% & 14.95\% \\
\gt 863.17 Hz & 7.59\% & 0.00\% & 7.59\% & 0.00\% & 34.18\% & 0.00\% & 50.63\% \\
\multicolumn{8}{c}{nonbarycentric birdies} \\
All & 6.02\% & 3.04\% & 8.76\% & 4.89\% & 7.81\% & 0.77\% & 68.71\% \\
\lt 0.11 Hz & 15.53\% & 12.42\% & 22.36\% & 15.53\% & 6.83\% & 4.35\% & 22.98\% \\
\gt 863.17 Hz & 0.00\% & 0.00\% & 0.00\% & 0.00\% & 5.10\% & 0.00\% & 94.90\% \\

\end{tabular}
\caption{Percent of birdies ranked highest by each scoring variant.\label{table:variants}}
\end{center}

\end{table}

Table \ref{table:variants} shows values of $F(V)$,
broken down by birdie type (barycentric or not)
and by the intrinsic bandwidth of the birdie
(only the smallest and largest ranges are shown).
From this data we can conclude that
a) the best score variant depends on birdie type and bandwidth;
b) each of the score variants is best
for a significant fraction of birdies.
Hence, in selecting multiplets for manual evaluation,
we examined the top-scoring multiplets
in all seven score variants.

\subsection{Defining event and candidate sensitivity}

Radio SETI projects typically consist of
a front end that finds {\em detections},
and a back end that finds a relatively small set of {\em candidates},
which are inspected manually and possibly reobserved.
In the case of SETI@home, candidates are high-scoring multiplets.

\begin{figure}[tbp]
\centerline{
\includegraphics[width=4cm]{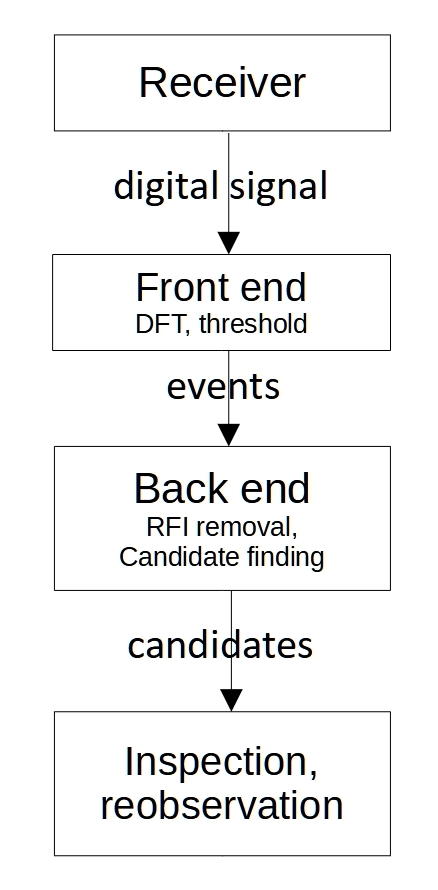}
}
\caption{The \change{general} structure of radio SETI projects.
\label{fig:zone2}}
\end{figure}

The sensitivity of the system is, \change{generally} speaking,
the minimum signal strength that it can detect.
But this must be defined \change{carefully,}
and it is important to distinguish between events and candidates.

Suppose a signal, with a flux $F$
and other parameters such as bandwidth $d\nu$,
arrives at the receiver at a particular time.
The signal may be detected, producing an {\em event}.
Whether it is detected is probabilistic because \change{it depends on}:
\begin{itemize}
\item noise from space (which varies with sky position);
\item noise in the receiver electronics;
\item RFI (which varies with time and frequency);
\item where the telescope was pointing relative to the signal source;
\item whether the telescope pointing was changing, and how fast;
\item sender and receiver acceleration, and corresponding Doppler drift.
\end{itemize}

We first define the system's {\em event sensitivity}, $S_{event}$.
To do this, we pick a probability threshold $P$ (say, 0.5).
$S_{event}$ is then defined as
the least value of $F$ for which the probability
of detecting the signal is at least $P$.
(We assume that the detection probability is a monotonic function of $F$:
the stronger the signal, the more likely it is to be detected).

$S_{event}$ depends on the signal bandwidth $d\nu$,
so it should be notated as $S_{event}(d\nu)$.
It also depends on the various factors listed above;
we can average these factors over the project's
observations to reduce it to a single number.

It is difficult to estimate a project's event sensitivity.
The sensitivity reported by most projects is a best-case value
that ignores most of the above factors.

But in any case, what matters for the purpose of detecting ET signals
is not event sensitivity,
but rather what we call {\em candidate sensitivity}.
Again, we pick a probability threshold $P$.
The candidate sensitivity $S_{candidate}$ is the least flux $F$
for which a signal with flux $F$
results in a candidate with probability at least $P$.

The $S_{candidate}$ and $S_{event}$ are related,
but they generally differ.
$S_{candidate}$ could be larger than $S_{event}$
for various reasons:
\begin{itemize}
\item In a noisy RFI environment
faint signals might be detected by the front end,
but almost all of them would be discarded as RFI
and hence would not produce candidates.

\item a faint signal might produce events,
but not enough to satisfy the requirements of the candidate detection algorithm.

\item RFI removal algorithms might be too aggressive
or the candidate detection algorithm might fail to find possible candidates.
\end{itemize}

Conversely, $S_{candidate}$ could be less than $S_{event}$.
Assume that the signals are persistent (always on)
and that the project observes sky positions
repeatedly or for long periods.
If a persistent signal is weaker than $S_{event}$,
it might be detected with below-threshold probability, say 0.4.
Hence, the signal could result in multiple events
which could be identified as a candidate by appropriate algorithms.\change{\footnote{It is important for SETI researchers to acknowledge 
this distinction.  A project
claiming a detection sensitivity of, for example, 10$^{-23}$ \wattpermsqr but only publishes and reobserves candidates with power $\gt 10^{-20}$ \wattpermsqr has a true sensitivity of $\sim10^{-20}$ \wattpermsqr.}}

Using the birdie mechanism,
we can estimate the candidate sensitivity of SETI@home.

\subsection{Candidate sensitivity for \change{continuous} signals
\label{subsection:sensitivity}}

For some birdies -- those with low power or whose positions
were not sufficiently close to any beam trajectory --
no spikes were generated.
Other birdies had spikes but no multiplets were produced
containing any of these spikes.
Of the birdies for which multiplets were produced,
we determined whether they were uncovered in the above sense.

We graphed, as a function of birdie power,
the fraction of birdies falling into these three classes:
those with spikes, those with multiplets,
and those with multiplets that were uncovered.
We did this separately for each of the 15 signal frequency
ranges, and for both barycentric and nonbarycentric signals.

\begin{figure}[tbp]
\centerline{
\includegraphics[scale=.5]{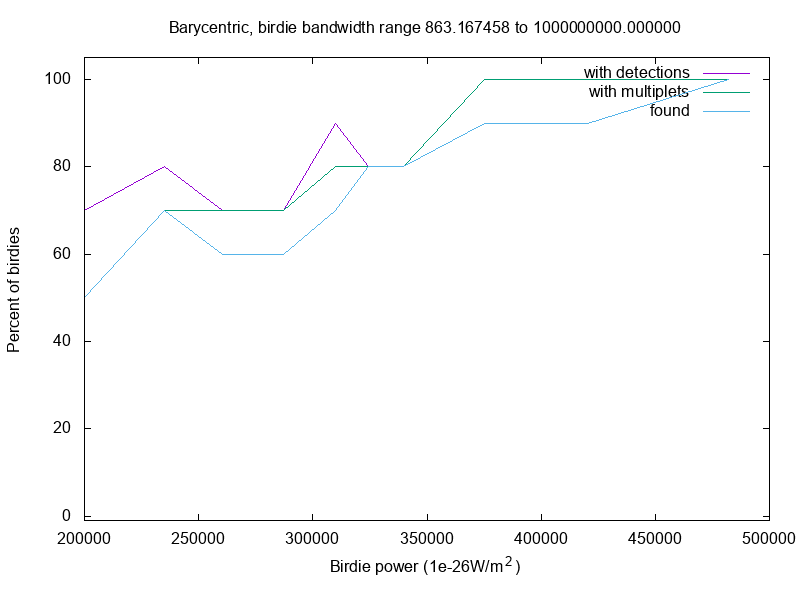}
}
\caption{Candidate sensitivity to barycentric signals with bandwidth $\ge$ 863 Hz
\label{figure:sens_bary_1}}
\end{figure}

\begin{figure}[tbp]
\centerline{
\includegraphics[scale=.5]{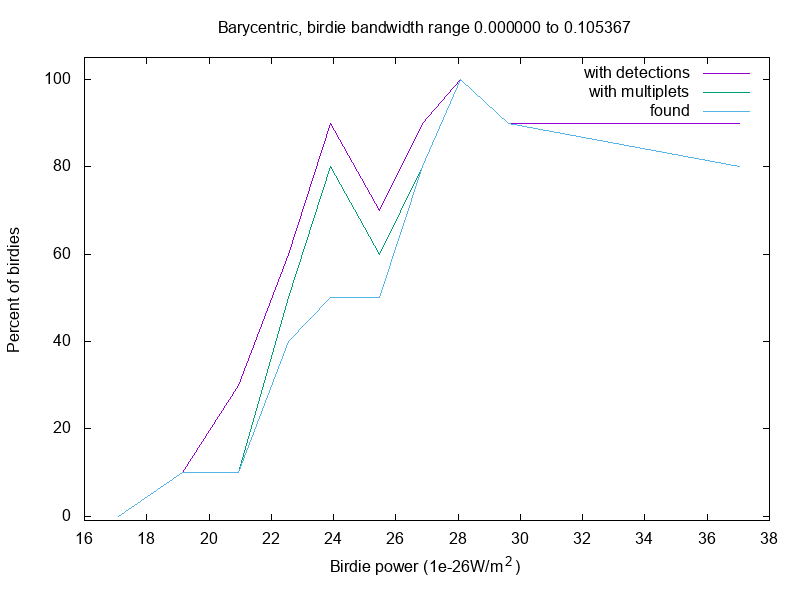}
}
\caption{Candidate sensitivity to barycentric signals with bandwidth $\le$ 0.1 Hz
\label{figure:sens_bary_3}}
\end{figure}

\begin{figure}[tbp]
\centerline{
\includegraphics[scale=.5]{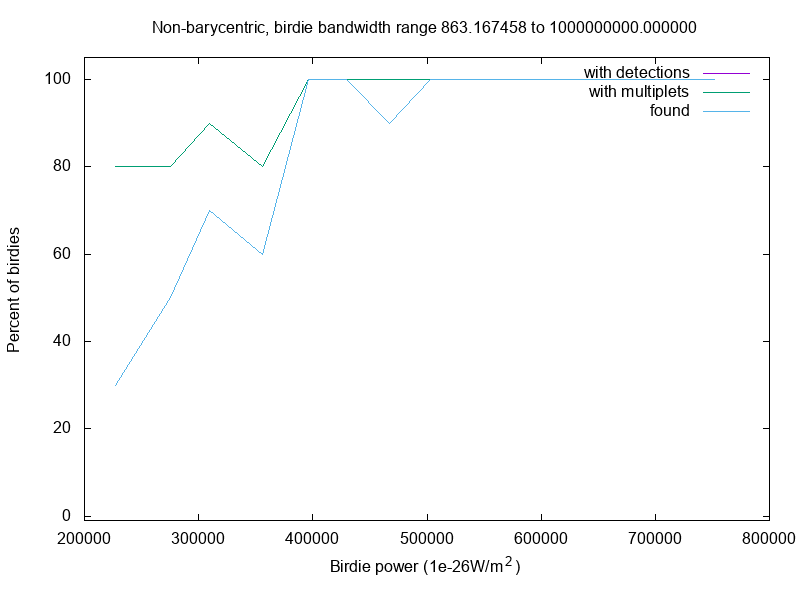}
}
\caption{Candidate sensitivity to nonbarycentric signals with bandwidth $\ge$ 863 Hz
\label{fig:sens_nonbary_1}}
\end{figure}

\begin{figure}[tbp]
\centerline{
\includegraphics[scale=.5]{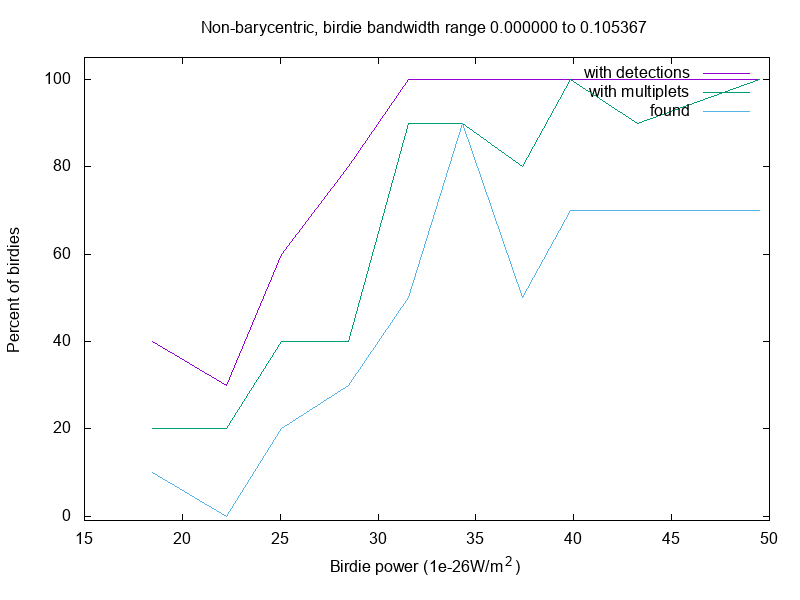}
}
\caption{Candidate sensitivity to nonbarycentric signals with bandwidth \lt 0.1 Hz
\label{fig:sens_nonbary_3}}
\end{figure}

In all cases, the fraction of
\change{birdies uncovered} increased with birdie power,
and above some power at least 80\% of birdies were uncovered.
This power is \change{an estimate of} our candidate sensitivity
to signals of that type.
Examples of this for barycentric birdies are shown
in Figures \ref{figure:sens_bary_1} to \ref{figure:sens_bary_3}.
We show the graphs for the 1st, 8th, and 15th bandwidth ranges;
the others are similar.
Analogous graphs for nonbarycentric birdies are shown
in Figures \ref{fig:sens_nonbary_1} to \ref{fig:sens_nonbary_3}.

Note \change{that} although birdies have the same range of SNR power,
the range of powers in terms of flux
varies widely with signal bandwidth.
This reflects the fact that our DFT frequency resolutions
are as small as 0.07 Hz, and we are much more sensitive
to signals with bandwidths in this range.

\begin{figure}[tbp]
\centerline{
\includegraphics[scale=.5]{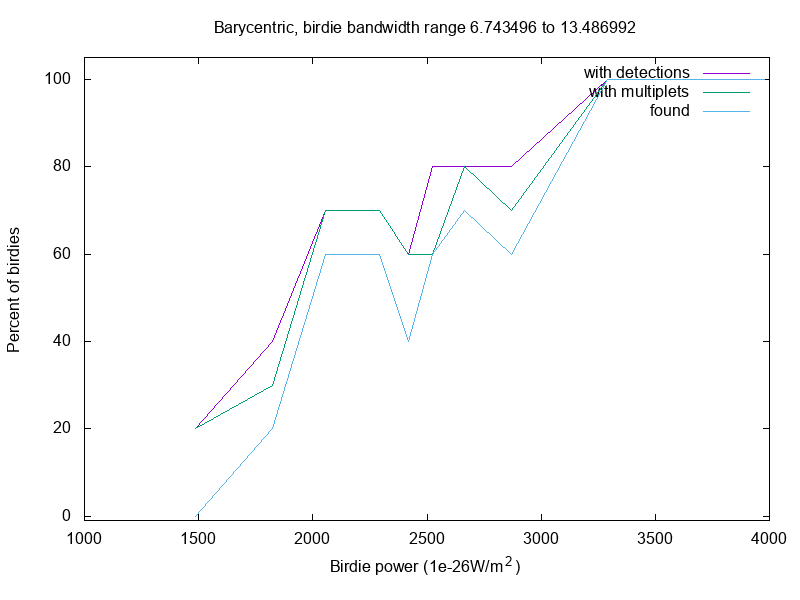}
}
\caption{Candidate sensitivity to barycentric signals with medium bandwidth (6 Hz to 13 Hz)}
\end{figure}

\begin{figure}[tbp]
\centerline{
\includegraphics[scale=.5]{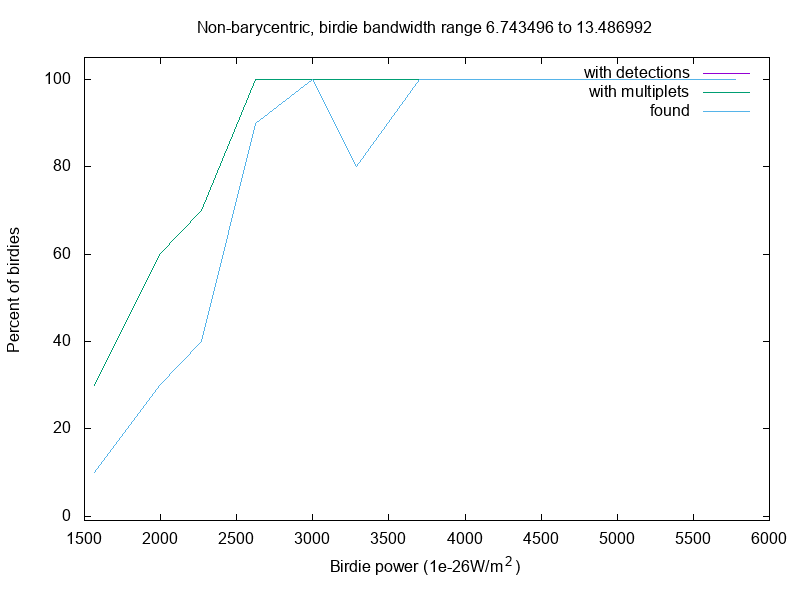}
}
\caption{Candidate sensitivity to nonbarycentric signals with bandwidth from 6 Hz to 13 Hz}
\end{figure}

These results are summarized in Table \ref{table:sensitivity}.
For each range of signal bandwidth
we show SETI@home's candidate sensitivity to barycentric
and nonbarycentric signals in that range
and the fraction of the sky in which we achieve this sensitivity.
As can be seen, SETI@home is most sensitive to
signals of bandwidths less than 0.1 Hz,
but achieves this sensitivity in only a small
fraction of the \change{sky ($\sim$2\%).}
This is because the telescope was moving too quickly
to detect such signals during most of our observations.

\begin{table}[tbp]
\begin{center}
\begin{tabular}{c c c c}
Signal& Candidate & Candidate & Sky \\
Bandwidth & Sensitivity (bary) & Sensitivity (nonbary) & Coverage \\
(\Hz) & ($10^{-26}\,\wattpermsqr$) & ($10^{-26}\,\wattpermsqr$) & (\%) \\
\hline
863--1726  & 320000 & 400000 & 100 \\
431--863  & 150000 & 150000 & 100 \\
215--431  & 90000 & 90000 & 100 \\
107--215  & 50000 & 45000 & 100 \\
53.9--107  & 21000 & 25000 & 100 \\
26.9--53.9  & 12000 & 12000 & 100 \\
13.4--26.9  & 5100 & 5100 & 99.73 \\
6.74--13.4  & 3300 & 2600 & 97.42 \\
3.37--6.74  & 1800 & 1500 & 93.51 \\
1.68--3.37  & 800 & 900 & 78.29 \\
0.842--1.68  & 350 & 650 & 50.67 \\
0.421--0.842  & 210 & 260 & 42.56 \\
0.210--0.421  & 120 & 160 & 21.43 \\
0.105--0.210  & 52 & 60 & 2.41 \\
0.052--0.105  & 28 & 40 & 2.24 \\
\end{tabular}
\caption{Summary of estimated candidate sensitivity and sky coverage\label{table:sensitivity}}
\end{center}
\end{table}

We conjectured that the total observing time in a birdie's pixel
might be correlated with the probability of finding the birdie,
but this turned out not to be the case.
The fraction of birdies \change{that were} uncovered did not consistently
exceed 80\% as this value increased.

\subsection{Selecting signal candidates for reobservation
\label{subsection:mp_select}}

The 300 meter Arecibo telescope collapsed in December 2020.
Only the Five-hundred-meter Aperture Spherical Telescope (FAST)
in China has sufficient sensitivity to check SETI@home candidates.
Together with our Chinese colleagues, we were granted
23 hours of dedicated observing time on FAST to
reobserve SETI@home candidates.
\change{Given the telescope time per reobservation
(see \S\ref{section:reobs})
this allows us to reobserve 92 candidates.}

We selected candidates \change{for reobservation} from among
the top-ranked multiplets in the various categories and score variants.
SETI@home is optimized for finding \change{continuous narrowband} signals.
In deciding how to budget our 92 reobservations
among multiplet categories,
we allocated more to the spike/Gaussian categories.
In total, we selected 70 spike/Gaussian candidates
(50 barycentric and 20 nonbarycentric),
12 pulse/triplet candidates,
and 10 autocorrelation candidates. 

Our algorithms for finding and ranking multiplets are not perfect;
manual evaluation is still needed.
For this purpose, we use the visualization tools described in
\S\ref{subsection:tools}.
These let us examine waterfall plots of the multiplet's detections
and see how detection times
are distributed over the observation intervals for the pixel.
A significant fraction of top-ranking multiplets
were identified as consisting of RFI and/or noise.

For each multiplet category and scoring variant,
the back-end software creates a list of multiplets
ordered by descending score.
Each multiplet has a summary page
showing the factors that went into the multiplet score,
information on each detection in the multiplet,
and rankings and comments by experts inspecting this multiplet and its detections.

\begin{figure}[tbp]
\centerline{
\includegraphics[width=18cm]{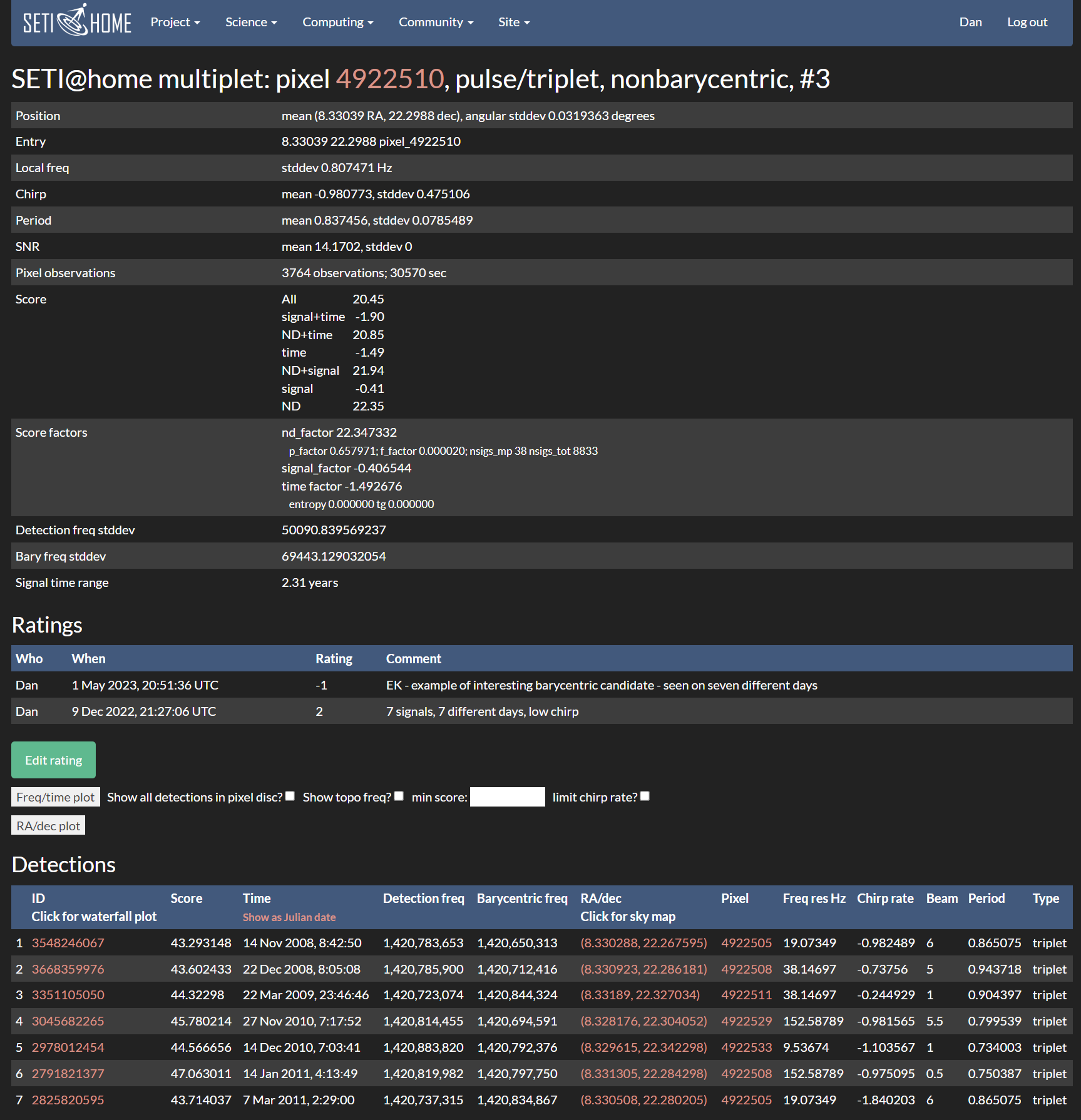}
}
\caption{Example of a multiplet summary page.
\label{fig:multipletsummarypage}}
\end{figure}

Figure~\ref{fig:multipletsummarypage} shows an example of a multiplet summary page.
This particular multiplet is a nonbarycentric pulse/triplet candidate
that is highly ranked manually and by algorithms because:
it consists of detections with high SNR from seven different observations
spanning 2.3 years at the same position on the sky (within 1 \change{beamwidth).}
These detection periods are similar (0.837 Hz),
the detections have a low negative drift rate (0.98 Hz/sec),
and their frequencies are within the span expected from a nonbarycentic
candidate (50 KHz RMS).
This multiplet was selected for reobservation at the FAST telescope.

\begin{figure}[tbp]
\centerline{
\includegraphics[width=18cm]{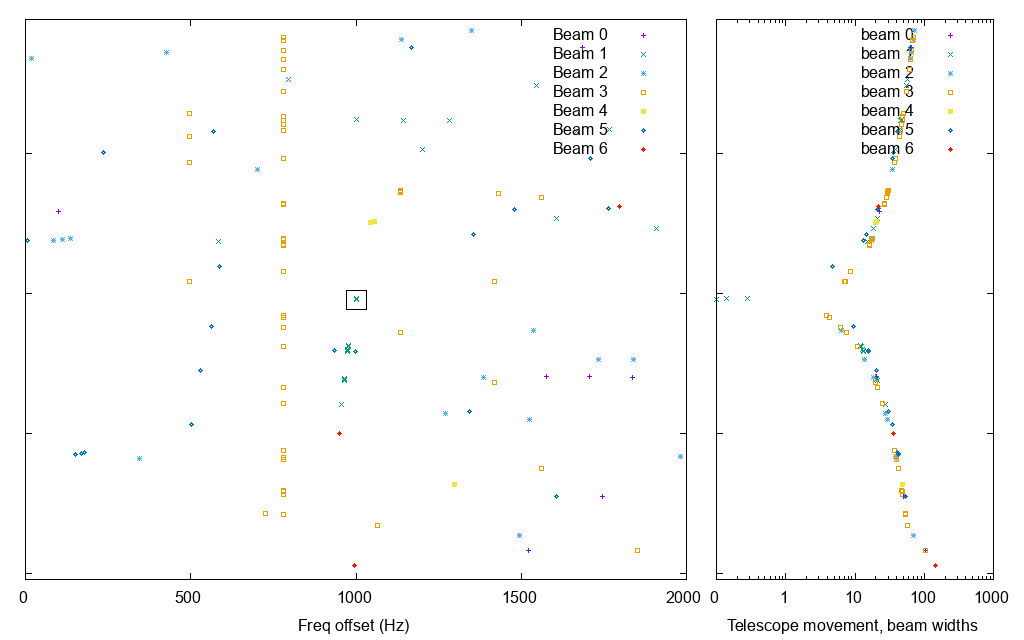}
}
\caption{\change{Waterfall plot showing detections from a high scoring multiplet that is likely due to RFI.  The detection at the center of the plot is indicated with a square.}
\label{fig:RFIexample}}
\end{figure}

Figure~\ref{fig:RFIexample} is a time vs. frequency waterfall plot from a multiplet
that was rejected manually because its detections are likely RFI
not \change{detected} by our RFI algorithms.
The left panel of the figure shows detections from different receiver beams.
The detection under examination is the blue X in the center, inside the small box.
The right panel shows the
relative telescope angle (in units of beam width) versus time.
The relative angle plotted is the difference between the telescope coordinates
(which change with time) and the coordinates of the detection being examined.
The detection in the center of the left plot is probably caused by RFI
because it was seen four other times at roughly the same frequency
while the telescope was pointing several beams away from the center detection.
(the four detections with blue X symbols sloping down and left from the center detection).

\begin{figure}[tpb]
\centerline{
\includegraphics[width=18cm]{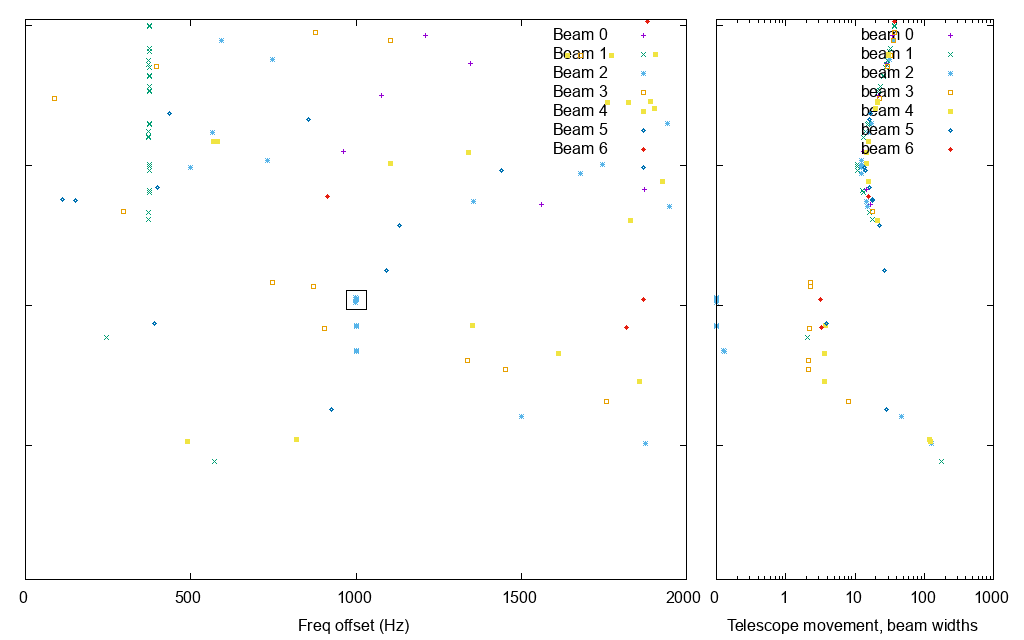}
}
\caption{Detections from a multiplet that was scored highly
by both algorithms and manually.
\label{fig:good_bary_multiplet_example}}
\end{figure}

In contrast, Figure~\ref{fig:good_bary_multiplet_example}
shows an example of a detection from a barycentric multiplet
that was given a high manual rating as well as a high machine rating.
The axes and symbols are similar \change{to those in} Figure~\ref{fig:RFIexample}.   
This multiplet was given high ratings for several reasons:
it was detected several different times during several different observations;
all these detections are clustered closely on the sky (within one \change{beamwidth)};
all detections are clustered tightly in barycentric frequency (50 Hz RMS);
none of the detections in the multiplet looked like RFI;
each time that sky position was observed,
there were several detections made while the telescope was tracking that sky position,
and when the telescope moved away from that position,
detections immediately ceased (this is consistent with a point source; it is unlikely RFI would do this,
and extremely unlikely that RFI would do this multiple times).
This multiplet was selected for reobservation at the FAST telescope.

For spike/Gaussian multiplets
we concentrated on the density and time score variants,
since that was most effective for finding both
barycentric and nonbarycentric birdies
(\S\ref{subsection:effectiveness}).
Many of the top-ranked multiplets consisted of
two detections within one minute.
We decided to rule \change{out these} and consider only multiplets
with a time span of at least 0.15 days.
We only selected spike/Gaussian barycentric multiplets
whose detections had drift rates consistent
with the barycentric reference frame (drift rates \lt 0.7 \hzpersec).
We first inspected highly ranked barycentric multiplets
containing three or more spikes or Gaussian signals;
we then considered highly ranked multiplets composed of 1 Gaussian and 1 spike,
and finally we selected a few multiplets with 2 spikes.


\section{Related work and contributions of SETI@home
\label{section:related}}

\change{As discussed in \S\ref{section:intro}, there have been many
radio SETI projects using both sky survey and targeted search methods.}
Like SETI@home, each of these projects has a front end and a back end.
Differences among front ends are discussed in \citep{instrument_paper}.
The differences among the back ends include the following.

\begin{description}

\item[Candidate birdies]
Some radio SETI projects (including SETI@home) tested
their event detection by injecting
constant-frequency or drifting sinusoids \citep{margot23}.
SETI@home is the first project to use high-level candidate birdies
that simulate persistent ET signals with a range of powers,
frequencies, bandwidths, and transmitter orbital parameters.
This lets us estimate the probability of detecting ET signals with different combinations of parameters
and provides a basis for evaluating and refining our
RFI rejection and candidate detection algorithms.

\item[RFI \change{detection} algorithms]

SETI@home's RFI detection algorithms
are novel to varying extents and may be useful in future
radio astronomy projects.
In particular, the algorithms for detecting zone and
drifting \change{narrowband} RFI are new.
Several projects have used some form
of cross-beam RFI rejection \citep{korpela09b, parsons04, harp16, setiburst17}.

\item[Repeated observations separated in time]
SETI@home recorded a decade-long archive of detections.
During twelve years of \change{observation,}
SETI@home has observed most of the Arecibo sky several times.
The observations of a given sky position are typically
widely separated in time.
Compared to projects in which a given sky
position is observed only once,
this increases the likelihood of detecting these
types of \change{signals:}

a) Sporadic signals,
such as signals from low duty cycle transmitters
(for example, a transmitter on the surface of a planet
may be invisible 50\% of the time as the planet rotates).

b) Scintillated signals:
scintillation can cause a \change{narrowband} signal
to be amplified or attenuated when propagating through the interstellar medium,
on timescales from hours to months,
possibly causing the signal to go above or below detectable thresholds
\citep{cordes97}.

\item[Search for long-duration nonbarycentric signals]
SETI@home searches for
nonbarycentric signals which, due to the transmitter's planetary motion,
vary widely (\nonbarycentricwindow) over time in received frequency.
The companion SERENDIP IV project at AO also
searched for such signals \citep{cobb00}.

\end{description}

SETI@home's algorithms
and techniques may be useful in future radio SETI projects.
Its software is open source, so it can be
used and adapted by such projects.
The source code is available at https://sourceforge.net/projects/seti-science/.

The web interfaces described here are visible online at
https://setiathome.berkeley.edu/nebula/.
We maintained a public blog describing the evolution of the back end:

https://setiathome.berkeley.edu/forum\_forum.php?id=1511.

\section{Future work
\label{section:future}}

\subsection{Reobservation of signal candidates \label{section:reobs}}

We are currently reobserving the selected signal candidates
(see \S\ref{subsection:mp_select}) at the
Five-hundred-meter Aperture Spherical Telescope (FAST) in China.
For each \change{candidate, 
we do a single scan across its sky position at 0.7 times the sidereal rate,
recording data with all 19 beams of the FAST telescope.
This scan rate allows the SETI@home client to find all detection types,
including spikes and Gaussians at their longest DFT lengths.
The scan covers 0.47 degrees and lasts 172 seconds.
Including slewing time,
it takes about 15 minutes to reobserve each candidate.  

We analyze the resulting data in three ways.
First, we use the SERENDIP VI system,
described in \citet{fastseti}
which operates commensally at FAST,
to search for strong narrowband
signals near the candidate's frequency.
 
Second, we use the SERENDIP VI system to record baseband data. 
A polyphase filter bank in SERENDIP VI generates 32768 15.26-kHz wide
``coarse channels" from the 500 \MHz (1.0-1.5 \GHz) bandwidth of the FAST receivers.
The coarse channels that overlap the \fullband SETI@home band are recorded as 16 bit complex samples in baseband format and analyzed in this format to
avoid quantization losses. 
We generate SETI@home workunits from these,
analyze them using the SETI@home client program,
and manually check the output for detections that match the candidate parameters. 

We can use Eqn.\,13 of \citet{instrument_paper} to evaluate the sensitivity of this method.
Our targets can be up to 27\degrees from the FAST zenith.
The beam performance from \citet{jiang20} gives a worst case system temperature of 19\,K at the zenith and 24\,K at 27\degrees from the zenith, versus 29\,K for ALFA,
and a gain of 13.2\,\Kjy relative to ALFA's 8.6\Kjy.
Minimal quantization losses result in a minimum effective area of 28\,200 m$^2$, or 2.56$\times$ that of our Arecibo observations.
However, because of the increased workunit bandwidth, our channel bandwidth has increased by 1.56$\times$.
For spikes, triplets and autocorrelation, the integration time has decreased by the same factor.
Therefore overall sensitivity to these signal types is improved about 2.0$\times$ over our Arecibo observations.
Because we conduct our observations at 0.7$\times$ the sidereal rate in order to compensate for the narrower beamwidths at FAST, integration times for Gaussians and pulses are essentially unchanged, resulting in a sensitivity improvement of 2.5$\times$ over our Arecibo data. 
 
Third, we manually investigate the expected frequency bands using coherent
 Doppler correction at the barycentric drift rate.
 We generate logarithmically scaled waterfall images at multiple time and frequency resolutions,
and look for features missed by the SETI@home client.
 
We began to reobserve candidates at FAST on September 24, 2022.
So far we have reobserved 80 candidates.
}
 
\subsection{Possible refinements of the front end}

As described \change{by} \citet{instrument_paper},
there are several ways in which the SETI@home front end
could be improved;
we could have workunits include data from all 14
beams and polarizations instead of just one.
We could merge the spike and Gaussian detection types.
These changes would improve event sensitivity.
This would also allow us to do cross-beam RFI rejection
in the client,
which would effectively improve \change{event }sensitivity
by letting us lower detection thresholds.

We have archived the ALFA data that we recorded,
so we could re-analyze this data with an improved
front end.
However, it would be preferable to do a new sky survey
using a telescope such as FAST or SKA,
for two reasons:
\begin{itemize}
\item
Since the start of SETI@home network bandwidth to the home
and PC storage capacity have increased greatly,
making it feasible to have much larger workunits.
\change{Thus,} it would be feasible to cover a much larger
frequency range: perhaps a factor of 10 or 100.
\item
The archived ALFA data was recorded via commensal
observation,
and its pointing trajectory was not \change{well suited}
to detecting \change{continuous} signals.
Future sky surveys should use, at least in part,
a pointing strategy optimized for continuous signal detection
(see \S\ref{subsec:strategy}).
\end{itemize}
\subsection{Possible refinements of the back end}

As described in \S\ref{section:birdies},
our use of birdies has served several purposes.
We currently use only \change{continuous} birdies,
and we generate only spike detections for them.
Generating Gaussians as well could improve our sensitivity estimates.
We could also generate pulsed birdies, and generate
pulses and triplets for them.
This would help us improve our RFI algorithms for pulsed \change{signals}
and would \change{allow us to} estimate our sensitivity to them.

Our multiplet-finding algorithms
(Sections \ref{subsection:nonbary_multiplet}
and \ref{subsection:bary_multiplet})
consist of a sequence of stages,
each of which removes detections to satisfy a constraint.
This approach may not lead to the \change{highest-scoring}
multiplets; for example, one stage may select detections that
must be removed by a later stage.
It might be better to use an algorithm,
perhaps based on clustering or other AI methods,
that combines constraints with score maximization.

The global drift-rate/frequency consistency constraint
for nonbarycentric multiplets (\S\ref{subsubsection:global_nonbary})
ensures that detections do not change too fast
for our planetary motion limits,
but it does not ensure that they match a frequency
trajectory resulting from a set of orbital
and rotational parameters within the limits.
A more sophisticated algorithm could do this.

It would also be possible to search the SETI@home detection archive for signals co-located (and co-accelerating) with Solar System objects such as planets, moons, asteroids, and Kuiper Belt \change{Objects (KBO)}.
\change{Such an} analysis would use object ephemerides for position clustering,
with frequency correction based on the changing object frames.

\subsection{\label{subsec:strategy}Pointing strategies for radio SETI sky surveys}

Perhaps the most important lesson from SETI@home is that,
in a sky survey with high frequency resolution, pointing matters.
Sensitivity to \change{continuous narrowband}
signals requires long observations
during which the telescope is drifting slowly or not at all.
Commensal observation may not provide \change{such} data.

We can estimate how much observing time is needed to survey the sky
with high frequency resolution.
As an example: in SETI@home there are 15M observable pixels.
Observing each for \maxtimebin (corresponding to a frequency resolution of \minbin) would take a total of 6.37 years.
A multibeam receiver can observe multiple pixels simultaneously;
so, for example, with the FAST 19-beam receiver the observing
time could in principle be reduced to 0.33 years.
(Ideally, for the reasons given in \S\ref{section:related},
this survey would be done at least twice).

What are the optimal pointing strategies?
This question \change{merits} study;
there is a trade-off between sensitivity and telescope time.
If we do a continuous sweep,
the beam crossing time should be at least
the bin duration of the longest DFT length.
And if it is longer -- say 2$\times$ or 3$\times$ -- this would provide
the RFI rejection that is built into the Gaussian fitting.
Looking at it the other way: if a constant slew rate is given,
we can avoid needlessly long DFT lengths.

For \change{RFI detection purposes,}
it is useful to observe each point after a short delay
(a minute or so).
This can be done with a back-and-forth pointing pattern.
With multibeam receivers,
a similar effect can be achieved by rotating the receiver
so that 2 or more beams lie in the direction of motion.
\change{However, in both cases,} the benefit comes at
the cost of increased observing time.

If a sky survey's telescope movement were regular --
for example, if beams always moved at the same rate --
the algorithms for some types of RFI detections,
for signal detection, and for candidate evaluation
could be more simple and probably more effective than
the flexible but complex algorithms we developed for SETI@home.
However, in light of the expense of large radio telescopes
and the high demand for telescope time from
other areas of radio astronomy,
sensitive sky surveys may be limited to commensal observing
for the foreseeable future.

\section{Acknowledgements}

Millions of SETI@home volunteers supplied computing power
for front-end data processing.
Volunteers contributed in many other ways,
such as providing technical support to other volunteers,
moderating message boards,
translating \change{Web site} text,
and porting the SETI@home application to GPUs;
see \cite{instrument_paper}.
Hundreds of people have helped develop
SETI@home's software and hardware systems;
the contributions of
Matt Lebofsky and Charlie Fenton were especially valuable.
David Gedye had the idea for SETI@home and assembled its initial team.

SETI@home has been supported by grants from Starwave, The Planetary Society,
the state of California,  National Science Foundation grant 1407804, 
the Marilyn and Watson Alberts SETI Chair fund,
and by donations from individuals.
We received equipment donations from Sun Microsystems,
Intel, NetApp, NVIDIA, AMD/Xilinx, Hewlett Packard,
 Fujitsu, Quantum, Seagate, Western Digital, and Packet Clearing House.

SETI@home's back-end processing used the Atlas cluster
at the Albert Einstein Institute in Hanover Germany.
\change{We thank} Bruce Allen for making this available.

This research used resources \change{from} the National Energy Research
Scientific Computing Center, a DOE Office of Science User Facility
supported by the Office of Science of the \change{US} Department of Energy
under Contract No. DE-AC02-05CH11231.

\bibliographystyle{aasjournal}
\bibliography{sah}

\appendix
\section{Algorithm: calculate per-pixel observation intervals\label{appendix:obs}}
\setcounter{figure}{0}
\renewcommand\thefigure{\thesection.\arabic{figure}}

\textbf{Input:} The pointing history of each beam $b$,
represented as a time-ordered list of (time, position) pairs.

\textbf{Output}: For each pixel $P$, a list $I(P)$ of
non-adjacent time intervals during which some beam's sky position was within $\AObeamwidth$ of the center of $P$.

The algorithm loops over beams $b$, calculating for each pixel $P$
a list $I(P, b)$ of
non-adjacent time intervals during which the position of $b$
was within $\AObeamwidth$ of the center of $P$.

Since the pointing sampling interval is 1 Hz, we assume that
if consecutive pointings are separated by at least 10 seconds,
there is a gap in the data.

We first add interpolated virtual pointings so
that no pixels are skipped.
To do this, we scan the list of $b$'s pointings in time order.
Let $A_1$ and $A_2$ be consecutive pointings.
If the time difference between $A_1$ and $A_2$ is more than 10 seconds,
we skip the pair.
Otherwise, for a \change{point $A$,} let $D(A)$ be the set of pixels
whose centers are within $\AObeamwidth$ of $A$.
$D(A)$ approximates the sky area seen by $b$ (in the above sense)
at the time of the pointing.
If $D(A_1)$ and $D(A_2)$ differ, we add a virtual pointing $A_{mid}$
halfway between $A_1$ and $A_2$ in time, \change{RA} and dec.
We then recursively do the same check for $(A_1, A_{mid})$ and $(A_{mid}, A_2)$, possibly adding further pointings.

Having added the interpolated pointings,
we compute, for each pixel $P$, a list $I(P, b)$ of observation intervals,
initially empty.
We scan the interpolated pointing list in time order,
again skipping pairs separated by more than 10 seconds.
For the other consecutive pairs $(A_1, A_2)$,
we add the time interval to $I(P, b)$ for each pixel $P$
in the union of $D(A_1)$ and $D(A_2)$.


After completing this scan, we merge adjacent intervals.
The resulting lists $I(P, b)$ are the output of the function.

Having \change{calculated} $I(P, b)$ for the seven beams $b$,
we merge these to form $I(P)$.

\section{Algorithm: generate birdie detections\label{appendix:birdies}}
\setcounter{figure}{0}
\renewcommand\thefigure{\thesection.\arabic{figure}}

\textbf{Input:} a set $S$ of birdies.

\textbf{Output:} a set of spike detections,
approximating what the SETI@home front end would have produced
had the birdie signals been present in its input data.

\begin{figure}[tb]
\centerline{\includegraphics[scale=.4]{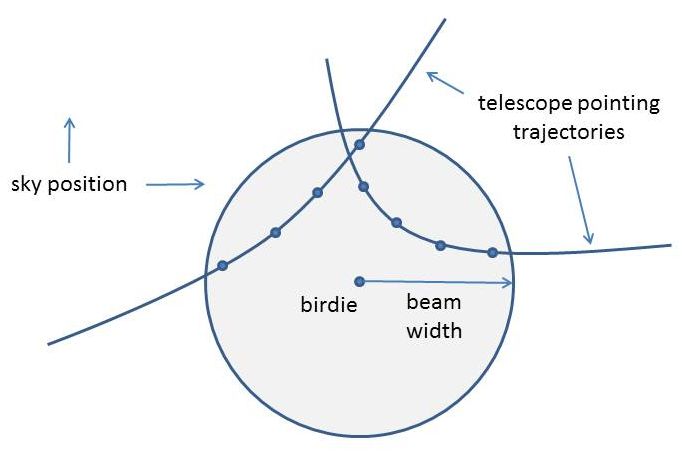}}
\caption{Birdie detections are generated when a telescope beam passes close to the birdie sky position.
\label{fig:birdie_gen}}
\end{figure}

For each of the 7 receiver beams $b$, we scan through its pointing history:
its sky positions over the SETI@home observing period,
interpolated as described in \S\ref{subsection:observation}.
\change{See Figure \ref{fig:birdie_gen}}

At each pointing $(P_{time}, P_{pos})$,
we find the set $S_0 \subseteq S$ of birdies $B$ with position close enough to $P_{pos}$ (within two beam widths)
that could potentially produce a detection.
We consider the time interval $I$ from $P_{time}$ to the time of the next pointing.
We loop over each DFT length $\fftlen$ as the birdie could potentially produce detections at any of these.
For each DFT length, we
loop over the DFT time intervals that overlap $I$.

For each birdie $B \in S_0$ we compute the angle $\theta$ between $P_{pos}$ and $\Bpos$.
Since the sensitivity of the beam is a Gaussian function,
the attenuation due to this angular separation is
\begin{equation}
A_{angle} = e^{-\frac{\theta}{\AObeamwidth}}
\end{equation}

Recall that $\fftlenres$ is the frequency resolution that corresponds to $\fftlen$.
The attenuation due to the disparity between $\fftlenres$ and $\Bbandwidth$ is
\begin{equation}
A_{freq} = \sqrt{\frac{\fftlenres}{\Bbandwidth}}
\end{equation}
if $\Bbandwidth > \fftlenres$, and
\begin{equation}
A_{freq} = \sqrt{\frac{\Bbandwidth}{\fftlenres}}
\end{equation}
otherwise.

To model the effects of noise, we add a random number $R$ from a normal distribution with mean 0 and stddev 1.

The resulting total power is

\begin{equation}
P = (B_{power} A_{angle} A_{freq})+ R
\end{equation}
If $P$ is above the threshold for spikes,
we create a spike $D$ with $\Dsnr = P$,
$\Dpos = P_{pos}$,
and $\Dtime$ equal to the midpoint of the DFT time interval.

$\Dnubary$ is
$\Bfreq$ plus the Doppler \change{shifts resulting from the movements of the sender and receiver.}
If $B$ is barycentric, the former is zero;
otherwise it is the Doppler shift due to \change{the motion of $B^\prime$} at time $t$.
The latter is $\AOdoppler(b, t)$.

$\Dnutopo$ is $\Dnubary + \AOdoppler(b, t)$.
We then round $\Dnutopo$ down
to a multiple of the frequency resolution of DFT length $\fftlen$.

$\Dchirp$ is the sum of sender and receiver components.
The former is the derivative of $B$'s Doppler shift
as determined by its planetary motion parameters,
or zero for barycentric birdies.
The latter is $\AOchirp(b, t)$.

Detections of a given DFT length $\fftlen$ are spaced in time by at least $\fftlendur$,
the duration corresponding to $\fftlen$.
So we must ensure that this holds for birdie detections.
\change{For long DFT lengths t}his may require skipping over pointing intervals
For short DFT lengths, we may create several detections within a single pointing interval.

\section{Algorithm: detect frequency-zone RFI\label{appendix:zone}}
\setcounter{figure}{0}
\renewcommand\thefigure{\thesection.\arabic{figure}}

\change{The prevalence of zone RFI varies with detection type $T$ and DFT length $\fftlen$,
so we process each combination separately.
The band fraction removed, denoted $\zonefrac$,
depends on the combination.

For each detection type and DFT length,
we divide the set of detections into 0.1-day {\em time windows}
and, for each window,
we flag frequency bins that have a statistical excess of detections.
Then, for each frequency bin,
we count the number of windows during which the bin was flagged.
We find the count threshold for which at least $\zonefrac$ of the bins are over threshold,
and flag detections in these bins as RFI.}

\textbf{Input:} A set $S$ of detections of a given type $T$ (spike or Gaussian) and DFT length $\fftlen$.

\textbf{Output:} A set $S^\prime \subseteq S$ of detections marked as RFI.

Algorithm parameters:

\begin{description}
\item[$\zonebinsize$] The \change{size of the frequency bin.}
This is twice the DFT frequency resolution,
to account for frequency imprecision due to Doppler drift.
\item[$\zonewindow$] The duration of a window: 0.1 days.
\item[$\zoneprob$] A probability threshold used in
deciding whether a bin has an excess of detections: $10^{-7}$.
\item[$\zonefrac$] The fraction of bins to flag as RFI, 
determined separately for each combination of
detection type and DFT length, as described above.
\end{description}
We divide the overall frequency range into bins of size $\zonebinsize$
Let $nbins$ be the number of bins.

We scan $S$ in time order, accumulating a window $W$ of detections.
$W$ is complete when
\begin{itemize}
\item adding the next detection to $W$ would cause its
time span to exceed $\zonewindow$, and
\item the number of detections in $W$
is at least half the average number of detections
of the given type and DFT length during a period of duration
$\zonewindow$,
averaged over the entire SETI@home observation period.
\end{itemize}

As we process windows, we maintain an array $FC[i]$
contain\change{ing}, for each frequency bin $i$,
the count of windows during which the bin had an excess of detections.

When a window is complete, we analyze it
to identify bins with a statistical excess of detections.
An RFI source may not be contained in a single bin;
it may \change{have a} significant bandwidth, or its frequency may vary.
So, after analyzing the bins singly,
we combine them into groups of 2, 4 and so on.
If one of these combined bins has an excess of detections,
all of its component bins are flagged.
This is described in the pseudocode in Fig.~\ref{fig:rfi_zone_alg}.

\begin{figure}[tbh]
\noindent\rule{\textwidth}{1pt}
	\begin{algorithmic}
	    \State $stride = 1$
	    \State $nsigs = |W|$
	    \For {$i=0; i < nbins; i \pluseq 1$}
	        \State $C[i] =$number of detections in bin $i$
	    \EndFor
	    \While{true}
	    \State $thresh = stride * \zoneprob/nbins$
	    \State $avg = stride * nsigs / nbins$
        \State $n$ = the smallest number for which $\Gamma(n, avg) < thresh$
		\For {$i=0; i < nbins; i \pluseq stride$}
		    \If {$C[i]$ is greater then $n$}
		        \State $F[i]...F[i+stride-1]$ = True
		        \State $C[i]$ = 0
		    \EndIf
		\EndFor
		\If {$stride*2 > nbins/2$}
		    \State break
		\EndIf
		\State $stride = stride * 2$
		\For {$i=0; i < nbins; i \pluseq stride$}
		    \State $C[i] = \sum( C[i] .. C[i+stride-1])$
		\EndFor
		\EndWhile
		\For {$i=0; i < nbins; i \pluseq stride$}
		    \If{$F[i]$}
		        \State $FC[i] \pluseq 1$
		    \EndIf
		\EndFor
	\end{algorithmic}
\noindent\rule{\textwidth}{1pt}
\caption{Frequency-zone RFI Algorithm\label{fig:rfi_zone_alg}}
\end{figure}

Having processed all windows,
the array $FC[i]$ contains the number of windows
in which bin $i$ was flagged as having an excess of detections.
Let $X$ be the $1-\zonefrac$ quantile of $FC$;
i.e. the value for which $\zonefrac$ of the bins are above $X$.

Let $B$ be the set of bins $i$ for which $FC[i] > X$.
Let $S^\prime$ be the subset of $S$ consisting
of detections whose bin is in $B$;
these detections are flagged as RFI.

\begin{figure}[tbp]
\centerline{
\includegraphics[width=12cm]{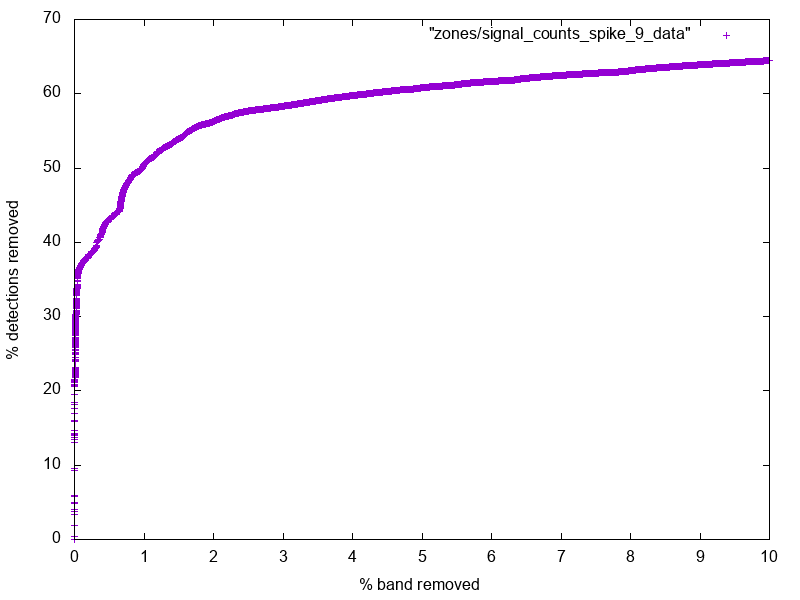}
}
\caption{Spikes removed as a function of frequency band removed
for DFT length 4096.  The knee of this curve (in this case about 2\%) determines the fraction of the frequency band removed by the zone RFI algorithm.
\label{fig:zone1}}
\end{figure}

\change{To determine appropriate values for $\zonefrac$,
we examined graphs of the fraction of signals removed
as a function of the fraction of frequency zones removed,
assuming that zones are removed in decreasing order of $E(Z)$.
These graphs -- one per detection type and DFT length --
have a knee because a small fraction of zones contain
disproportionate numbers of detections.
An example is shown in Figure \ref{fig:zone1}.
We choose $\zonefrac$ to be a value slightly above this knee.}

\change{Note that in deciding whether to remove a zone, we use window count instead of number of detections in the zone.
If we used the number of detections, a zone could be removed
due to a large excess of detections in a short time range.
We do not want to remove that zone at all times because
an ET signal could occur in that zone at a different time.
This RFI would be removed by other RFI filters
(such as the drifting filter; see below).}

\section{Algorithm: detect drifting RFI\label{appendix:drifting}}
\setcounter{figure}{0}
\renewcommand\thefigure{\thesection.\arabic{figure}}

\textbf{Input:} A set S of detections.

\textbf{Output:} A set $S^\prime \subseteq S$ of detections marked as drifting RFI.

Algorithm parameters:
\begin{description}
\item[$\driftdt$] The duration of a triangle: 600 s.

\item[$\driftmaxdrift$] The maximum drift rate: 5\,\hzpersec.

\item[$\driftminangle$] Count only detections at least this far from $D$: $\frac{3}{2} \AObeamwidth$.

\item[$\driftntriangles$] The number of triangles in the fan: 21.

\item[$\driftpa$] Probability threshold for single triangles (see below): $10^{-8}$.

\item[$\driftpb$] Probability threshold for opposed pairs of triangles: $10^{-4}$.
\end{description}

First, divide $S$ into clusters as described above.
For each cluster, select one representative detection.
Let $C$ denote the set of these representatives.

For a detection $D \in C$, consider the rectangle
in time/frequency space centered at $\Dtime$, $\Dnutopo$
and of duration $\driftdt$ and frequency width $\driftdt \driftmaxdrift$.
Construct $2\driftntriangles$ equal-area triangles
with one vertex at $D$ and the other two on
the top or bottom edge of the rectangle,
as shown in Figure \ref{fig:drifting}.

For each triangle $T$, compute the number $F(T) \subseteq C$ of detections
that a) lie within $T$ and b) whose angular distance from
$D$ is greater than $\driftminangle$.

For each triangle $T$, compute an
estimate $P(T)$ of the probability that $T$ contains at least $F(T)$ detections.
This is done as follows.
For each of the two triangle fans (before and after $D$),
let A and B be the median and mean, respectively,
of $F(T)$ over the triangles $T$ in the fan.
If A is greater than zero, let
\begin{equation}
M = \min(A, B)
\end{equation}
otherwise let
\begin{equation}
M = \min(1, B)
\end{equation}
For each triangle T in the fan, let
\begin{equation}
P(T) = \Gamma(M, F(T))\\
\end{equation}
where $\Gamma$ is the incomplete Gamma function.
This estimates the probability of a triangle containing
at least $F(T)$ detections, given an average of $M$.

For each triangle $T$, consider the opposite triangle and
the two adjacent triangles.
If for $T$ and an opposing triangle $S$, $P(T)$ and $P(S)$ are
less than $\driftpb$,
then flag $D$ as RFI.
This tracks bands of drifting RFI whose drift rate changes over time.
In addition, if for any triangle $T$, $P(T)$ is less than $\driftpa$,
then flag $D$ as RFI.
This flags the start and end of bands of drifting RFI.

If $D$ is flagged as RFI by the above procedure,
also flag the detections in the same cluster as $D$.

Repeat the above procedure for each $D \in C$.

\section{Algorithm: detect medium-term pulse and triplet RFI\label{appendix:pulse_rfi}}
\setcounter{figure}{0}
\renewcommand\thefigure{\thesection.\arabic{figure}}

\textbf{Input}: A set $S$ of pulses or triplets.

\textbf{Output}: A subset $S^\prime \subseteq S$ of detections flagged as RFI.

Algorithm parameters:

$\periodwindow$.  The period over which we look for RFI features: 10 minutes.

$\periodprob$.  The probability threshold for a statistical excess: $10^{-3}$.

$\periodangle$.  Detections are considered far if
the angle between their positions exceeds
$\frac{3}{2} \AObeamwidth$.

Scan $S$ in time order, maintaining a window $W$ of duration at most $\periodwindow$.
When adding a detection would exceed $\periodwindow$,
process $W$ as follows:

Bin the detections in $W$ by frequency,
with bin size twice the median
bandwidth for detections of that type (38 Hz for pulses, 305 Hz for triplets).

For each bin $B_i$, find the detection $D_i$ whose time is closest to the midpoint of $W$.
Let $L_i$ be the set of detections in $B_i$ with time before $D_i$ that are
far from $D_i$,
that is, whose angular distance from $D_i$ is at least $\periodangle$.
Let $U_i$ be the set of far detections after $D_i$.

Let $N_{far}$ be the number of far signals in all bins.

Let $M$ be the median size of the $L_i$ and $U_i$.

Let $X$ be the least integer such that 
\begin{equation}
\Gamma(X, M) < \periodprob
\end{equation}
$X$ is the threshold for a statistical excess.

If, for a bin $i$, $|L_i|>X {\rm\ and\ }|U_i|>X$, then flag all
the detections in that bin as RFI, including those that are not far.

Having processed $W$, advance the window by $\periodwindow/2$
and continue scanning $S$.

\section{Algorithm: detect multi-beam RFI\label{appendix:mb}}
\setcounter{figure}{0}
\renewcommand\thefigure{\thesection.\arabic{figure}}

\textbf{Input:} A set $S$ of detections.

\textbf{Output:} A subset $S^\prime \subseteq S$ of detections flagged as RFI.

For each detection $D$, we define a rectangle $R(D)$
in time/frequency space.
The center of the rectangle is ($\Dtime$, $\Dnutopo$).
The size in each dimension represents the uncertainty in that attribute.
If another detection $D_2$ is contained in this rectangle,
then $D_2$ probably has the same source (cosmic or RFI) as $D$.

The size of $R(D)$ in the frequency dimension is $\fftlenres$,
the frequency resolution corresponding to
the detection's DFT length $\fftlen$.
The size in the time dimension is $\Ddur$.

For each detection $D^\prime \in S$,
we find the set $T$ of detections $D$ of the same type
for which ($\Dtime$, $\Dnutopo$) lies in $R(D^\prime)$.
If $D$ is a pulse or triplet, we include only detections
with periods close to $\Dperiod$.

We then count the number $N$ of detections $D^\prime \in T$
for which the angle between $\Dpos$ and $\Dposprime$
exceeds $1.75\AObeamwidth$.
These detections are unlikely to have the same source as $D$.
If $N$ is at least half the size of $T$,
we flag $D$ as RFI.

\section{Algorithm: local drift-rate/frequency pruning\label{appendix:local_drift}}
\setcounter{figure}{0}
\renewcommand\thefigure{\thesection.\arabic{figure}}

\textbf{Input:} A set $S$ of detections with a time span of at most $\mdtlocal$.

\textbf{Output:} A subset $S^\prime \subseteq S$ that satisfies
the local drift-rate/frequency consistency constraints.

\begin{figure}[tbp]
\centerline{\includegraphics[scale=.6]{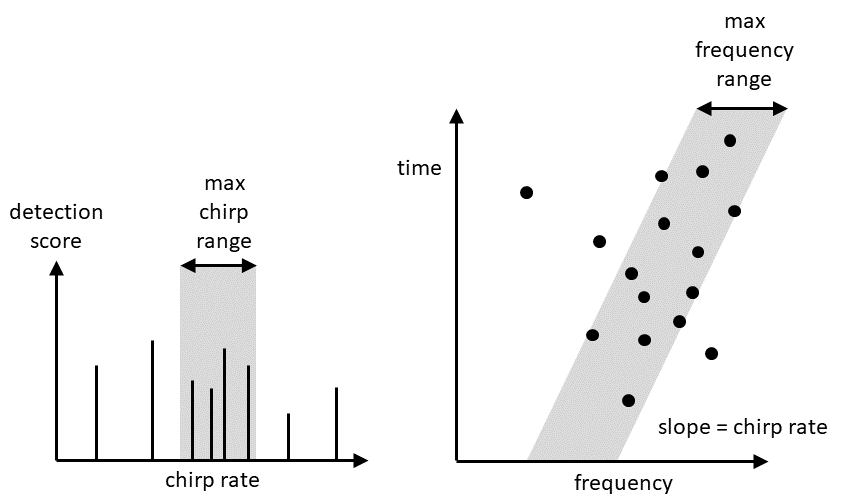}}
\caption{Local drift-rate/frequency pruning.
We a) find the drift rate interval containing the most detection power
and b) find the frequency interval which
when time-shifted by $C_{median}$ contains the most detection power.
\label{fig:prune}}
\end{figure}

We enforce the constraints
by pruning detections that violate them.
First, we find a range of drift rates
that contains a concentration of detections.
To do this, we scan $S$ in order of drift rate
and find the range of drift rates [$C_0$..$C_1$]
of size $\mdclocal$
in which the sum of detection scores is greatest.
We then discard detections with drift rates outside the interval.
\change{See Figure~\ref{fig:prune}.}

Let $C_{median}$ be the median drift rate of the remaining detections.
For each detection $D$,
we compute the barycentric frequency adjusted for the
sender component of $C_{median}$:
\begin{align}
{\Dnuadj} = & {\Dnubary} + C_{median}  (\Dtime - T_0))\\
  & - (\Dnubary - \Dnutopo) \nonumber
\end{align}
where $T_0$ is the time of the earliest detection in $S$.

We then scan the detections in order of increasing
$\Dnuadj$,
and find the interval [$f_0$..$f_1$] of width $\mdfnonbary$
in which the sum of the detection scores is greatest.
$S^\prime$ is then the set of detections $D$ for which $\Dnuadj$ lies in that interval.

\section{Algorithm: global drift-rate/frequency pruning\label{appendix:global_drift}}
\setcounter{figure}{0}
\renewcommand\thefigure{\thesection.\arabic{figure}}

\textbf{Input:} A set $S$ of detections.

\textbf{Output:} A subset $S^\prime\subseteq S$ satisfying both the local and global drift-rate/frequency constraints.

We scan $M$'s detections in time order,
using the above algorithm to process windows of duration at most $\mdtlocal$.
This produces, for each window,
a locally consistent set of detections $W_i$.

We maintain a list $W_1...W_n$ of windows that we have already processed
and which satisfy the global constraint.
For each such window $W$, we store its set of detections,
the median time, drift rate, and frequency of these detections,
and the sum of their scores;
the latter approximates the contribution of $W$
to the overall multiplet score.

\begin{figure}[tbp]
\centerline{\includegraphics[scale=.5]{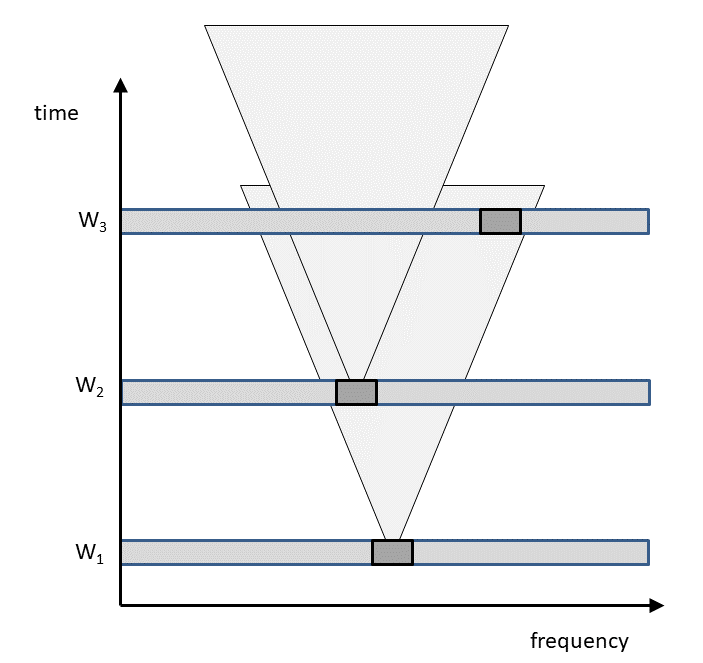}}
\caption{Global drift-rate/frequency pruning.  The detections selected in window $W_3$ are not compatible with those selected in $W_2$; we must
discard the one with lower value.
\label{fig:chirp}}
\end{figure}

We say that two windows are consistent if their
median times, drift rates, and frequencies satisfy the
constraint in \S\ref{subsubsection:global_nonbary}.
Having processed a window $W_{n+1}$, we scan the list $W_1...W_n$
and find the set of windows that is inconsistent with $W_{n+1}$.
If the sum of the values of these windows is less than the value of $W_{n+1}$,
we discard the conflicting windows and add $W_{n+1}$ to the window list.
Otherwise, we remove the detections we selected from $W_{n+1}$,
rerun the local consistency algorithm with the remaining detections
in that time window,
and check for consistency with the other windows.
This is repeated until no detections remain, in which case we discard $W_{n+1}$.
\change{see Figure~\ref{fig:chirp}.}

\section{Algorithm: Find a barycentric multiplet in a frequency band\label{appendix:find_bary_freq}}
\setcounter{figure}{0}
\renewcommand\thefigure{\thesection.\arabic{figure}}

\textbf{Input:} A set $S$ of detections in
a pixel disk and a frequency band of $\mdfbary$ Hz.

\textbf{Output:} A multiplet $M$, \change{consisting} of detections in $S$,
that satisfies \change{the} multiplet constraints,
or null if no such $M$ is found.

We first remove detections whose
drift rate is outside the limits described
in \S\ref{subsection:constraints}.
We then do time-overlap pruning (see \S\ref{subsubsection:timeoverlap})
on the remaining detections.
Finally, for spike/Gaussian multiplets,
we enforce frequency-variation consistency
(see \S\ref{subsubsection:freqvar})
by scanning the remaining detections in time order
and removing groups of detections that violate the constraint.
  
If at least two detections remain, return $M$.
Otherwise, return null.

\section{Algorithm: Find barycentric multiplets in a detection disk\label{appendix:find_bary_disk}}
\setcounter{figure}{0}
\renewcommand\thefigure{\thesection.\arabic{figure}}

\textbf{Input:} A set S of detections (the detection disk of a pixel).

\textbf{Output:} A set of barycentric multiplets made up of detections in S.

We scan through the detections in $S$ in order of increasing $\Dnubary$,
maintaining a window $W$ whose range of $\Dnubary$ is at most $\mdfbary$.
We skip detections that do not satisfy the
drift rate constraint for barycentric multiplets
(see \S\ref{subsubsection:chirp_constraints}).

When adding a detection to $W$ would exceed $\mdfbary$,
we look for a multiplet within $W$,
using the algorithms described in Section
\ref{subsection:bary_multiplet}.
If this produces a multiplet $M$,
we advance the window beyond the maximum frequency of $M$ and continue.
Otherwise (if no multiplet is found in $W$)
we append the next detection to $W$ and remove
detections from the start of $W$ as needed
to limit its frequency range to $\mdfbary$.

\begin{figure}[tbp]
\centerline{\includegraphics[scale=.5]{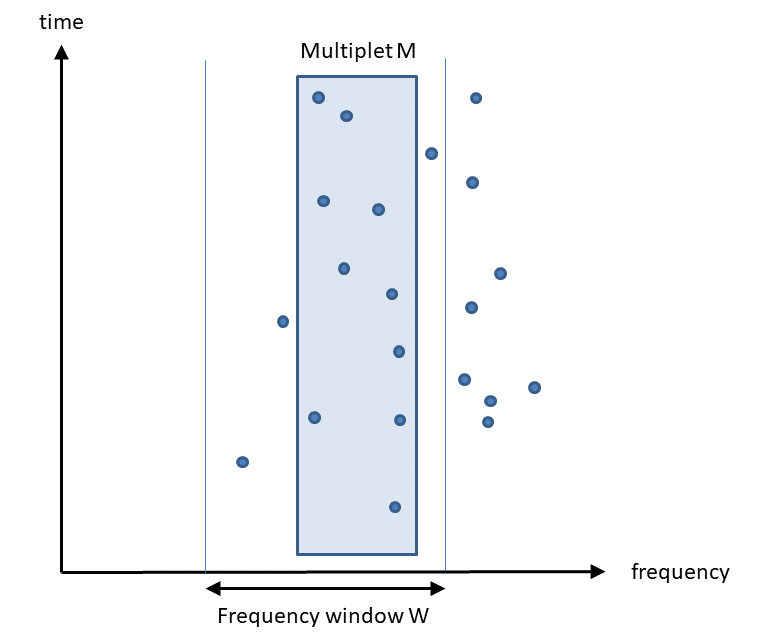}}
\caption{$M$ is the highest scoring multiplet in $W$,
but there may be a higher scoring multiplet that overlaps $M$
using higher-frequency detections.
\label{fig:multiplet}}
\end{figure}

When we identify a multiplet $M$ in a window $W$,
we do not immediately output it
because there may be a multiplet $M_2$,
including detections with frequencies above $W$,
that overlaps $M$ in frequency and has a higher score;
see Figure~\ref{fig:multiplet}.

In this case, we want to output $M_2$ rather than $M$.
To do this, we maintain a {\em reserved multiplet} $M_R$ that is possibly null.
When we identify a new multiplet $M$
and $M_R$ is not null:
if the frequency range of $M$ is completely above that of $M_R$,
we output $M_R$ and reserve $M$.
If the frequency ranges overlap and
$M_R$ has a higher score than $M$,
we output $M_R$ and advance the window beyond its maximum frequency;
otherwise, we reserve $M$.

If we find a multiple $M$ and there is no reserved multiplet,
let $A$ and $B$ denote the minimum and maximum frequencies of $M$.
If the next detection to be scanned has a frequency of at most $A + \mdfbary$,
we reserve $M$ and move the window $W$ to start at $A$
(that is, remove detections from $W$ with frequencies less than $A$).
Otherwise, we output $M$, and move $W$ to start at $B$.

\end{document}